\documentclass[10pt,letterpaper]{article}
\usepackage{opex3}
\usepackage{color}
\usepackage{amsmath}
\usepackage{amsfonts}
\usepackage{amssymb}
\usepackage{bm}
\usepackage{textcomp}
\usepackage{empheq}
\usepackage{cite} % hyphenated consecutive citations
\usepackage[colorlinks=true,linkcolor=blue,citecolor=blue,urlcolor=blue]{hyperref}
\renewcommand{\Re}{\operatorname{Re}}
\renewcommand{\Im}{\operatorname{Im}}

\newcommand{\citeasnoun}[1]{Ref.~\cite{#1}}

\renewcommand{\eqref}[1]{Eq.~(\ref{eq:#1})}
\newcommand{\eqreftwo}[2]{Eqs.~(\ref{eq:#1},\ref{eq:#2})}

\newcommand{\Eqref}[1]{Equation~(\ref{eq:#1})}
\newcommand{\figref}[1]{Fig.~\ref{fig:#1}}
\newcommand{\Figref}[1]{Figure~\ref{fig:#1}}
\newcommand{\cc}[1]{#1^*}
\newcommand{\secref}[1]{Sec.~\ref{sec:#1}}
\newcommand{\appref}[1]{App.~\ref{app:#1}}
\newcommand{\vect}[1]{\mathbf{#1}}
\newcommand*{\tens}[1]{\overline{\overline{#1}}}
\newcommand{\BigO}[1]{\ensuremath{\operatorname{O}\bigl(#1\bigr)}}

\newcommand*\widefbox[1]{\fbox{\hspace{2em}#1\hspace{2em}}}

\begin{document}

%%%%%%%%%%%%%%%%%% title page information %%%%%%%%%%%%%%%%%%
\title{Fundamental limits to optical response in absorptive systems}

\author{Owen D. Miller$,^{1,*}$ Athanasios G. Polimeridis,$^2$ M. T. Homer Reid,$^1$ Chia Wei Hsu,$^3$ Brendan G. DeLacy,$^4$ John D. Joannopoulos,$^5$ Marin Solja\v{c}i\'{c},$^5$ and Steven G. Johnson$^1$}
\address{$^1$Department of Mathematics, Massachusetts Institute of Technology, Cambridge, MA, 02139, USA\\
$^2$Skolkovo Institute of Science and Technology, 143025 Moscow Region, Russia\\
$^3$Department of Applied Physics, Yale University, New Haven, CT, 06520, USA\\
$^4$U.S. Army Edgewood Chemical Biological Center, Research and Technology Directorate, Aberdeen Proving Ground, MD, 21010, USA\\
$^5$Department of Physics, Massachusetts Institute of Technology, Cambridge, MA, 02139, USA}

\email{$^*$odmiller@math.mit.edu} %% email address is required

%%%%%%%%%%%%%%%%%%% abstract and OCIS codes %%%%%%%%%%%%%%%%
\begin{abstract}
    At visible and infrared frequencies, metals show tantalizing promise for strong subwavelength resonances, but material loss typically dampens the response. We derive fundamental limits to the optical response of absorptive systems, bounding the largest enhancements possible given intrinsic material losses. Through basic conservation-of-energy principles, we derive geometry-independent limits to per-volume absorption and scattering rates, and to local-density-of-states enhancements that represent the power radiated or expended by a dipole near a material body. We provide examples of structures that approach our absorption and scattering limits at any frequency; by contrast, we find that common ``antenna'' structures fall far short of our radiative LDOS bounds, suggesting the possibility for significant further improvement. Underlying the limits is a simple metric, $|\chi|^2 / \Im \chi$ for a material with susceptibility $\chi$, that enables broad technological evaluation of lossy materials across optical frequencies.
\end{abstract}

 % REPLACE WITH CORRECT OCIS CODES FOR YOUR ARTICLE, MINIMUM OF TWO; Avoid using the OCIS codes for “General” or “General science” whenever possible.

%%%%%%%%%%%%%%%%%%%%%%% References %%%%%%%%%%%%%%%%%%%%%%%%%
%\bibliographystyle{osajnl}
%\bibliography{/home/odmiller/texmf/bibtex/bib/library,fl_paper_bibitems}

%%%%%%%%%%%%%%%%%%%%%%%%%%  Body  %%%%%%%%%%%%%%%%%%%%%%%%%%
\section{Introduction}
\label{sec:intro}
At optical frequencies, metals present a tradeoff: their conduction electrons enable highly subwavelength resonances, but at the expense of potentially significant electron-scattering 
losses~\cite{Ozbay2006,Maier2007,Boltasseva2011,Atwater2010,Naik2011,Tassin2012,Arnold2009,Khurgin2010,Raman2013,Khurgin2015}. In this article we formalize the tradeoff between resonant enhancement and loss, deriving limits to the absorption within, the scattering by, and the local density of states (LDOS)~\cite{Novotny2012,Joulain2003,Martin1998,DAguanno2004,OskooiJo13-sources,Liang2013a} near a lossy, absorptive body of arbitrary shape. Given a material of susceptibility $\chi(\omega)$, the limits depend only on a material enhancement factor $|\chi(\omega)|^2 / \Im \chi(\omega)$ and on the incident-beam energy density (leading to a potential $1/d^3$ LDOS enhancement for a metal--emitter separation $d$). The power scattered or dissipated by a material body must be smaller than the total power it extracts from an incident beam; we show that this statement of energy conservation yields limits to the magnitudes of the internal fields and polarization currents that control the scattering properties of a body. Unlike previous bounds~\cite{Purcell1969,Sohl2007,Sohl2007a,Gustafsson2007,Pozar2009,Liberal2014,Wheeler1947,Chu1948,McLean1996,Wang2006a,Tassin2012,Raman2013,Hugonin2015}, our limits do not depend on shape, size, or topology, nor do they diverge for zero bandwidth. The crucial ingredient is that our bounds depend on $\chi$ and are finite only for realistic lossy materials---for idealized lossless materials such as perfect conductors, arbitrarily large optical responses are possible. We provide examples of structures that approach the theoretical limits, and also specific frequency ranges at which common structures fall far short. Our bounds apply to any absorptive system, and thus provide benchmarks for the response of metals, synthetic plasmonic materials (doped semiconductors)~\cite{West2010,Boltasseva2011,Naik2011,Naik2013,Law2013}, and surface-phonon-polariton materials across visible and infrared wavelengths, resolving a fundamental question~\cite{Ozbay2006,Maier2007,Boltasseva2011,Atwater2010,Naik2011,Tassin2012,Khurgin2010,Arnold2009,Raman2013,Khurgin2015} about the extent to which resonant enhancement can overcome intrinsic dissipation.

There has been intense interest in exploiting ``plasmonic''~\cite{Maier2007} effects, which arise for materials with permittivities that have negative real parts, in metals at optical frequencies. Geometries incorporating such materials are capable of supporting highly subwavelength surface resonances~\cite{Ozbay2006,Maier2007}. Yet such a material has inherent loss arising from the typically significant imaginary part of $\chi$. Even for applications in which absorption is the goal, material loss dampens resonant excitations and reduces the overall response. This tradeoff between resonant enhancement and absorption has been investigated for specific geometries amenable to semianalytical methods, leading to a variety of geometry-dependent material dependences. For example, in the quasistatic limit, coated spheres absorb energy at a rate proportional to $|\chi|/\Im \chi$~\cite{Averitt1999}, whereas spheroids absorb energy at a rate~\cite{Bohren1983} proportional to $|\chi|^2 / \Im \chi$. Surface modes at planar metal--insulator interfaces exhibit propagation lengths proportional to $\left(\Re \varepsilon\right)^2 / \Im \varepsilon$ at very low frequencies~\cite{Raether1988} (in the Sommerfeld--Zenneck regime~\cite{Maier2007}), but near their surface-plasmon frequencies their propagation lengths are approximately proportional to $\sqrt{\Im \varepsilon}$ (cf. \appref{plasmon_prop_length}). In electron energy loss spectroscopy (EELS), the electron scattering cross-section is proportional to a ``loss function'' $\Im (-1/\varepsilon) = (\Im \varepsilon) / |\varepsilon|^2$ that enables experimental measurement of bulk plasmon frequencies~\cite{Pines1964,Keast2005}.

We show that $|\chi|^2 / \Im \chi$ is a universal criterion for evaluating the optimal response of a metal, with a suitable generalization in \secref{limits} for more general media that may be anisotropic, magnetic, chiral, or inhomogeneous. For most materials, $|\chi|^2 / \Im \chi$ increases as a function of wavelength (as demonstrated in \secref{metal_limits}), suggesting that if an optimal structure is known, the plasmonic response of a metal can potentially be much greater \emph{away} from its bulk- and surface-plasmon frequencies. For effective-medium metamaterials, our bounds apply to both the underlying material parameters as well as to the effective medium parameters, with the \emph{smaller} bound of the two controlling the maximum response. Thus effective-medium approaches cannot circumvent the bounds arising from their constitutive materials; however, they may find practical application if they can achieve resonances at frequencies that are otherwise difficult to acheive with the individual materials.

The limits derived here arise from basic energy considerations. An incident field $\vect{E}_{\rm inc}$ interacting with a scatterer generates polarization currents $\vect{P}$ that depend on both the incident field and on the shape and susceptibility of the body. A lossy scatterer dissipates energy at a rate proportional to the squared magnitude of the currents, $|\vect{P}|^2$. At the same time, the total power extracted from the incident beam, i.e. the ``extinction'' (absorption plus scattering), is proportional to the imaginary part of the overlap integral of the polarization currents with the incident field, $\sim \int_V \cc{\vect{E}_{\rm inc}} \cdot \vect{P}$, which is known as the electromagnetic optical theorem~\cite{Lytle2005,Newton1976,Bohren1983,Jackson1999} and can be understood physically as the work done by the incident field to drive the induced currents. The overlap integral is only linear in $\vect{P}$ whereas the absorption depends quadratically on $\vect{P}$. If the magnitude of $\vect{P}$ could increase without bound, then, the absorption would become greater than extinction, resulting in a physically impossible negative scattered power. Instead, there is a limit to the magnitude of the polarization field, and therefore to the scattering properties of any body comprising the lossy material. We make this argument precise in \secref{limits}, where we employ variational calculus to derive general limits for a wide class of materials, and we also present limits specific to metals, which are typically homogeneous, isotropic, and nonmagnetic at optical frequencies. We consider here only bulk susceptibilities, excluding nonlocal or quantum effects~\cite{Scholl2012,Ciraci2012,Eggleston2015}. The key results are the limits to absorption and scattering in Eqs.~(\ref{eq:optScatSigma},\ref{eq:optAbsSigma},\ref{eq:optScatUMetal}--\ref{eq:optAbsSigmaMetal}) and the limits to LDOS enhancement in \eqreftwo{rhoRadLimitMetal}{rhoNrLimitMetal}. Before deriving the limits, we present the volume-integral expressions for absorption, scattering, and radiative and nonradiative LDOS in \secref{AbsScatLDOS}. 

In \secref{structures} we compare the response of a number of structures towards achieving the various limits. For far-field absorption and scattering, we find that ellipsoidal nanoparticles are ideal and can reach the limits across a wide range of frequencies. For near-field enhancement of power expended by a dipole into radiation or absorption, we find that it is much more difficult to reach the limits. The nonradiative LDOS near a planar metal surface reaches the limit at the ``surface-plasmon frequency'' of the metal. At lower frequencies, common structures (thin films, metamaterials) fall far short. At all frequencies, common designs for radiative LDOS enhancement exhibit suboptimal response. These results suggest the possibility for significant design improvement if the limits are achievable (i.e. ``tight'').

Previous limits to the electromagnetic response of metals have emphasized a variety of limiting factors. At RF and millimeter-wave frequencies, where the response can be bounded relative to that of a perfect electric conductor (PEC)~\cite{Hansen2006}, the Wheeler--Chu--McLean limit~\cite{Wheeler1947,Chu1948,McLean1996,Hansen2006} bounds the radiative $Q$ factor of an electrically small antenna. At optical frequencies, absorption loss increases and often dominates relative to radiative loss. There are known lower bounds on the absorptive $Q$ for low-loss, quasistatic structures~\cite{Wang2006a}, and more generally for metals of any size with susceptibilities comprising Lorentz--Drude oscillator terms~\cite{Raman2013}.  

Limits to frequency-integrated extinction are also known. Purcell derived the first such limit, using the Kramers-Kronig relations~\cite{Jackson1999} to bound the integrated response of spheroidal particles to their electrostatic ($\omega=0$) induced dipole moments~\cite{Purcell1969}. Recently the limits have been extended to arbitrary shapes~\cite{Sohl2007,Sohl2007a,Gustafsson2007}, but one obtains a different limit for each shape. Moreover, it is important in many applications to disentangle the single-frequency response and the bandwidth, and to do so separately for absorption and scattering.

Single-frequency absorption and scattering limits have primarily been derived via spherical-harmonic decompositions, originally for spherically symmetric scatterers~\cite{Hamam2007,Ruan2011} and later for generic ones~\cite{Pozar2009,Liberal2014}. This approach has been generalized recently, yielding limits in terms of the inverse of a scattered-field-operator~\cite{Hugonin2015}, although it the inverse of such an operator is seemingly difficult to bound without resorting to spherical harmonics. In \secref{limits} we show that the scattered-field-operator approach and our material-dissipation approach share a common origin in volume-integral equations. The key distinction is that the scattered-field operator and its corresponding limits are independent of material but dependent on structure, whereas our limits incorporate material properties and are \emph{independent} of structure. Both classes of limits apply to any linear body. In \secref{ext_disc} we provide a more detailed comparison, finding that the spherical-harmonic limits may provide better design criteria at lower (e.g. rf) frequencies, whereas our limits should guide design at higher frequencies, especially in the field of plasmonics.

This work was partly inspired by our recent bounds~\cite{Miller2014} on extinction by quasistatic nanostructures. In that work we derived bounds via sum rules of quasistatic surface-integral operators~\cite{Fuchs1975,Fuchs1976}; equivalently, we could have~\cite{MiltonPC} derived the bounds via analogous constraints in composite theory~\cite{Milton1981,Bergman1981}. The key distinction between this work and our previous work~\cite{Miller2014} is that here we find limits in the full Maxwell regime, such that our bounds apply to any structure at any size scale, and they apply to functions of the scattered fields (e.g. scattered power and radiative LDOS), which have zero amplitude in quasistatic electromagnetism. An additional benefit of our simplified energy-conservation approach is that we can bound the responses of anisotropic, magnetic, and/or inhomogeneous media, whereas the surface-integral sum-rule approach only works for isotropic and nonmagnetic materials. In this work we also consider the local density of states, which we did not consider previously and which represents an important design application. 

\section{Absorption, scattering, and LDOS expressions}
\label{sec:AbsScatLDOS}
\begin{figure}
\centering
\includegraphics[width=0.7\linewidth]{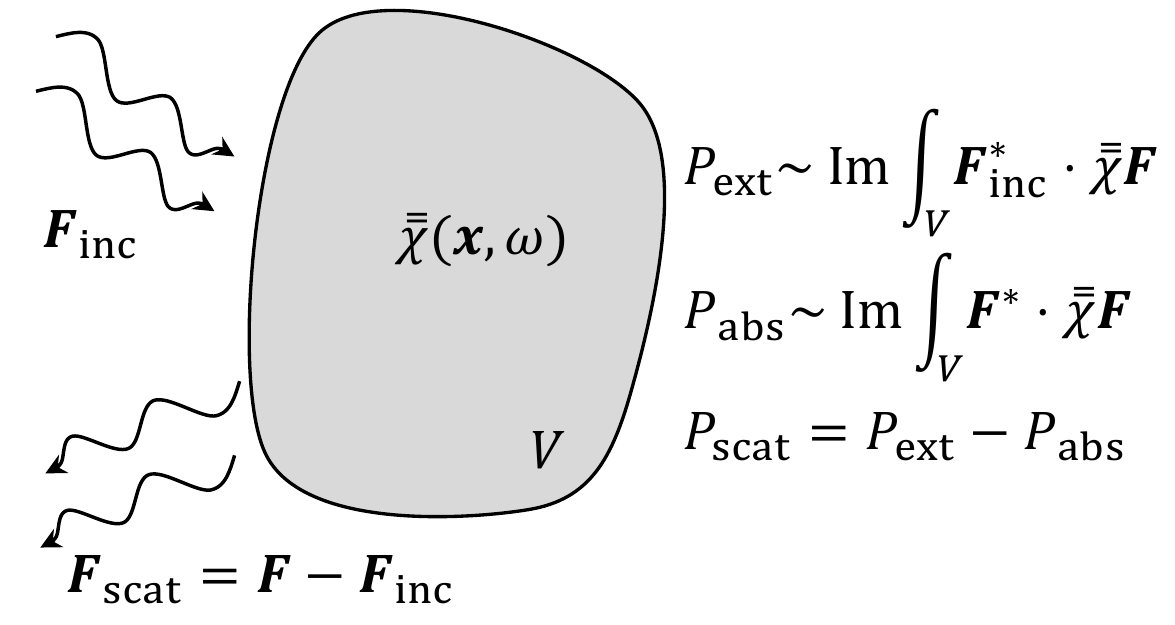}
\caption[Schematic illustration of scattering problem]{\label{fig:scattering_problem}Scattering problem under consideration. An incident field $\vect{F}_{\rm inc} = \begin{pmatrix}\vect{E}_{\rm inc} & Z_0 \vect{H}_{\rm inc}\end{pmatrix}^T$ impinges on a lossy scatterer with a susceptibility tensor $\tens{\chi}(\vect{x},\omega)$. The shape and topology of the scatterer are arbitrary: It may be periodic, extend to infinity, or consist of multiple particles. The limits presented in \secref{limits} hinge on the fact that absorption is a quadratic functional of the electromagnetic fields, whereas extinction, by the optical theorem, is the imaginary part of a linear functional of the fields. In \secref{limits}, we present general limits for tensor susceptibilities and also simplified limits for metals.}
\end{figure}
We consider lossy media interacting with electromagnetic fields incident from fixed external sources (e.g. plane waves or dipole sources). \Figref{scattering_problem} illustrates the conceptual setup: a generic scatterer with susceptibility tensor $\tens{\chi}$ absorbs, scatters, and extinguishes (extinction defined as absorbed plus scattered power) incident radiation at rates proportional to volume integrals over the scatterer. In this section we present the known volume-integral expressions for absorption and scattering, and we also derive volume-integral expressions for the power expended by a dipole near such a scatterer, which is either radiated to the far field or absorbed in the near field. Relative to a dipole in free space, the enhancement in power expended is given by the relative increase in the local density of states (LDOS)~\cite{OskooiJo13-sources}.

The scatterer is taken to consist of a lossy, local material that is possibly inhomogeneous, electric, magnetic, anisotropic, or bianostropic (chiral). We assume the scatterer is in vacuum, with permittivity $\varepsilon_0$, permeability $\mu_0$, impedance $Z_0 = \sqrt{\mu_0/\varepsilon_0}$, and speed of light $c = 1 / \sqrt{\varepsilon_0 \mu_0}$. Extending the limits of the following section to non-vacuum and possibily inhomogeneous backgrounds is relatively straightforward and is discussed in \secref{ext_disc}. The response of the scatterer can be described by induced electric and magnetic polarization currents, $\vect{P}(\vect{x})$ and $\vect{M}(\vect{x})$, which satisfy the constitutive field relations
\begin{equation}
\begin{aligned}
    \vect{D}(\vect{x}) &= \varepsilon_0 \vect{E}(\vect{x}) + \vect{P}(\vect{x}) \\
    \vect{B}(\vect{x}) &= \mu_0 \left[\vect{H}(\vect{x}) + \vect{M}(\vect{x})\right].
\end{aligned}
\end{equation}
For the general class of materials considered here, the currents $\vect{P}$ and $\vect{M}$ each depend on both $\vect{E}$ and $\vect{H}$ through a unitless 6$\times$6 susceptibility tensor $\tens{\chi}$~\cite{OskooiJo13-sources,Chew1995}:
\begin{align}
    \begin{pmatrix}
        \vect{P} \\
        \frac{1}{c} \vect{M}
    \end{pmatrix}
    = \varepsilon_0 \tens{\chi}
    \begin{pmatrix}
        \vect{E} \\
        Z_0 \vect{H}
    \end{pmatrix}
    = \varepsilon_0 \tens{\chi} \vect{F}
    \label{eq:vol_currents}
\end{align}
where $\vect{F}$ is a generalized vector field containing both electric and magnetic fields. For isotropic media with relative permittivity $\varepsilon_r$ and relative permeability $\mu_r$, the susceptibility tensor comprises only the diagonal elements $\varepsilon_r - 1$ and $\mu_r - 1$. Lossy media as considered here have susceptibilities that satisfy the positive-definiteness condition (for an $e^{-i\omega t}$ time convention)~\cite{Chew1995,Welters2014}
\begin{align}
    \omega \Im \tens{\chi} > 0,
    \label{eq:lossyDefinition}
\end{align}
where $\Im \tens{\chi} = (\tens{\chi} - \tens{\chi}^\dagger)/2i$ (and $^\dagger$ represents the conjugate transpose).

When light impinges on the scatterer, the absorbed, scattered, and extinguished powers can be written as overlap integrals of the internal currents and fields~\cite{Polimeridis2015}. The absorption (dissipation) within such a medium is the work done by the total fields on the induced currents, given by the expression~\cite{Kong1972}: 
\begin{align}
P_{\rm abs} = \frac{\varepsilon_0 \omega}{2} \Im \int_V \cc{\vect{F}}(\vect{x}) \cdot \tens{\chi}(\vect{x}) \vect{F}(\vect{x}) \,{\rm d}V
    \label{eq:pAbs}
\end{align}
where the asterisk denotes complex conjugation. \Eqref{pAbs} reduces to the usual $(\varepsilon_0\omega/2)\left(\Im \chi\right) \int_V \left|\vect{E}\right|^2$ for homogeneous, isotropic, nonmagnetic media. 

The total power extracted from the incident fields---the extinction---is the sum of the absorbed and scattered powers and can be computed by the optical theorem~\cite{Jackson1999,Born1999}. Although commonly written as an integral over fictitious effective surface currents~\cite{Jackson1999}, the optical theorem can also be written as a volume integral over the polarization currents~\cite{Lytle2005,Hashemi2012,Polimeridis2015}, representing the work done by the \emph{incident} fields on the induced currents:
\begin{align}
P_{\rm ext} = \frac{\varepsilon_0 \omega}{2} \Im \int_V \cc{\vect{F}}_{\rm inc}(\vect{x}) \cdot \tens{\chi}(\vect{x}) \vect{F}(\vect{x}) \,{\rm d}V
    \label{eq:pExt}
\end{align}
where, as for $\vect{F}$, we define $\vect{F}_{\rm inc}$ by
\begin{align}
    \vect{F}_{\rm inc} = \begin{pmatrix}
        \vect{E}_{\rm inc} \\
        Z_0 \vect{H}_{\rm inc}
    \end{pmatrix}.
\end{align}
The scattered power is the difference between extinction and absorption:
\begin{align}
    P_{\rm scat} &= P_{\rm ext} - P_{\rm abs} \nonumber \\
                 &= \frac{\varepsilon_0\omega}{2} \Im \int_V \left[\cc{\vect{F}}_{\rm inc}(\vect{x}) - \cc{\vect{F}}(\vect{x}) \right] \cdot \tens{\chi}(\vect{x}) \vect{F}(\vect{x}) \,{\rm d}V.
    \label{eq:pScat}
\end{align}

In addition to absorbing and scattering light, structured media can also alter the spontaneous emission rates of nearby emitters. Increased spontaneous emission shows exciting potential for surface-enhanced Ramam scattering (SERS)~\cite{Nie1997,Kneipp1997}, fluorescent imaging~\cite{VanZanten2009,Schermelleh2010}, thermophotovoltaics~\cite{Laroche2006,Basu2011}, and ultrafast light-emitting diodes (LEDs)~\cite{Lau2009}. The common metric for the enhanced emission rate is the (electric) local density of states (LDOS), which represents the density of modes weighted by the relative energy density of each mode's electric field at a given position~\cite{OskooiJo13-sources}. Equivalently, and more generally, the LDOS enhancement represents the enhancement in the total power expended by an electric dipole radiator~\cite{OskooiJo13-sources,Wijnands1997,Xu2000}, into either radiation or dissipation. Similar to extinction, the total LDOS is the imaginary part of a linear functional of the induced electric fields~\cite{Joulain2003}:
\begin{align}
    \rho_{\rm tot} &= \frac{1}{\pi \omega} \Im \sum_j \hat{\vect{s}}_j \cdot 
            \vect{E}_{s_j}(\vect{x}_0),
    \label{eq:rho_source_loc}
\end{align}
where $\vect{E}_{s_j}$ denotes the field from a dipole source at $\vect{x}_0$ polarized in the $\hat{\vect{s}}_j$ direction, with a dipole moment $\vect{p}_0 = \varepsilon_0 \hat{\vect{s}}_j$, and the sum over $j$ accounts for all possible orientations (the conventional LDOS corresponds to a randomly oriented dipole~\cite{Joulain2003}).

To connect the LDOS to the material properties, we rewrite it as a volume integral over the fields within the scatterer. The total field at the source position, $\vect{E}_{s_j}(\vect{x}_0)$, consists of an incident field and a scattered field:
\begin{align}
    \vect{E}_{s_j}(\vect{x}_0) = \vect{E}_{\textrm{inc},s_j}(\vect{x}_0) + \vect{E}_{\textrm{scat},s_j}(\vect{x}_0).
    \label{eq:inc_scat_dec}
\end{align}
The incident field is known---it is the field of a dipole in vacuum---and can be left as-is (note that the imaginary part of a dipole field does not diverge at the source location~\cite{Novotny2007}). The scattered field arises from interactions with the scatterer and is the composite field from the induced electric and magnetic currents, radiating as if in free space:
\begin{align}
    \vect{E}_{\rm scat}(\vect{x}_0) = \int_V \left[\tens{G^{EP}}(\vect{x}_0,\vect{x}) \vect{P}(\vect{x}) + \tens{G^{EM}}(\vect{x}_0,\vect{x}) \vect{M}(\vect{x}) \right]
    \label{eq:scat_int_PM}
\end{align}
where $\tens{G^{EP}}$ and $\tens{G^{EM}}$ are free-space dyadic Green's functions~\cite{Tai1994,Chew1995}. We could at this point insert \eqreftwo{inc_scat_dec}{scat_int_PM} into \eqref{rho_source_loc} and have a volume-integral equation for the total LDOS. However, note that the Green's functions in \eqref{scat_int_PM} represent the fields of free-space dipoles and the excitation in this case is also a dipole. We follow this intuition to replace the Green's functions by the incident fields.

Whereas the source dipole at $\vect{x}_0$ generates incident fields at points $\vect{x}$ within the scatterer, the Green's functions in \eqref{scat_int_PM} are the fields at $\vect{x}_0$ from a dipole at $\vect{x}$. By reciprocity~\cite{Rothwell2001} one can switch the source and destination points of vacuum Green's functions
\begin{subequations}
\begin{align}
    G^{EP}_{ij}(\vect{x}_0,\vect{x}) &= G^{EP}_{ji}(\vect{x},\vect{x}_0) \label{eq:ep_reciprocity} \\
    G^{EM}_{ij}(\vect{x}_0,\vect{x}) &= -\mu_0 G^{HP}_{ji}(\vect{x},\vect{x}_0) \label{eq:em_reciprocity}
\end{align}
\end{subequations}
where for clarity we indexed the Green's function tensors. Now one can see that the product $\hat{\vect{s}}_j \cdot \tens{G^{EP}} \vect{P}$ equals $\varepsilon_0^{-1} \vect{E}_{\rm inc} \cdot \vect{P}$. The magnetic Green's function yields the incident magnetic field, with a negative sign arising from reciprocity, as in \eqref{em_reciprocity}. \Eqref{inc_scat_dec} can be written
\begin{align}
    \hat{\vect{s}}_j \cdot \vect{E}_{s_j}(\vect{x}_0) = \hat{\vect{s}}_j \cdot \vect{E}_{\textrm{inc},s_j}(\vect{x}_0) + \int_V \vect{\widetilde{F}}_{\textrm{inc},s_j} \cdot \tens{\chi} \vect{F}_{s_j}
    \label{eq:rho_vol_int}
\end{align}
where we have defined
\begin{align}
    \vect{\widetilde{F}}_{\rm inc} = 
    \begin{pmatrix}
        \vect{E}_{\rm inc} \\ 
        -Z_0 \vect{H}_{\rm inc}
    \end{pmatrix}.
\end{align}
Inserting the field equation, \eqref{rho_vol_int}, into the LDOS equation, \eqref{rho_source_loc}, yields the total LDOS as the sum of free-space and scattered-field contributions:
\begin{align}
    \rho_{\rm tot} = \rho_0 + \frac{1}{\pi \omega} \Im \sum_j \int_V \vect{\widetilde{F}}_{\textrm{inc},s_j} \cdot \tens{\chi} \vect{F}_{s_j}.
\end{align}
where $\rho_0 = \omega^2 / 2\pi^2 c^3$ is the free-space electric LDOS~\cite{Joulain2005}. It is typically more useful to normalize $\rho$ to $\rho_0$:
\begin{align}
    \frac{\rho_{\rm tot}}{\rho_0} = 1 + \frac{2\pi}{k^3} \Im \sum_j \int_V \vect{\widetilde{F}}_{\textrm{inc},s_j} \cdot \tens{\chi} \vect{F}_{s_j}
    \label{eq:rhoTot}
\end{align}
where $k = \omega / c$. \Eqref{rhoTot} relates the total LDOS to a volume integral over the scatterer, which will enable us to find upper bounds to the response in the next section.

For many applications it is important to distinguish between the power radiated by the dipole into the far-field (where it may be imaged, for example) and the power absorbed in the near field (which may productively transfer heat, for example). Absorbed power is given by \eqref{pAbs}, and thus the nonradiative LDOS enhancement $\rho_{\rm nr}/\rho_0$ is given by \eqref{pAbs} divided by the power radiated by a dipole (of amplitude $\varepsilon_0$) in free space, $P_{\rm rad} = \varepsilon_0 \omega^4 / 12\pi c^3$~(\citeasnoun{Jackson1999}):
\begin{align}
    \frac{\rho_{\rm nr}}{\rho_0} = \frac{2\pi}{k^3} \Im \sum_j \int_V \cc{\vect{F}}_{s_j} \cdot \tens{\chi} \vect{F}_{s_j}.
    \label{eq:rhoNr}
\end{align}
Finally, just as the scattered power in \eqref{pScat} is the difference between extinction and absorption, the radiative part of the LDOS is the difference between the total and nonradiative parts:
\begin{align}
    \frac{\rho_{\rm rad}}{\rho_0} = 1 + \frac{2\pi}{k^3} \Im \sum_j \int_V \left[\vect{\widetilde{F}}_{\textrm{inc},s_j} - \cc{\vect{F}}_{s_j}\right] \cdot \tens{\chi} \vect{F}_{s_j}
    \label{eq:rhoRad}
\end{align}

\section{Limits}
\label{sec:limits}
Given the power and LDOS expressions of the previous section, upper bounds to each quantity can be derived by exploiting the energy conservation ideas discussed in the introduction (\secref{intro}). The extinction is the imaginary part of a linear function of the polarization currents, whereas absorption is proportional to their squared magnitude (and scattered power is the difference between the two), and thus energy conservation yields finite optimal polarization currents and fields for each quantity. 

Just as one can use gradients to find stationary points in finite-dimensional calculus, one can use variational derivatives~\cite{Gelfand2000} to find stationary points of a \emph{functional} (i.e. a function of a function). It is sufficient here to consider functionals of the type $P = \int \cc{f} g$, which arise in the power expressions, Eqs.~(\ref{eq:pAbs},\ref{eq:pScat},\ref{eq:rhoNr},\ref{eq:rhoRad}). The variational derivative of $P$ with respect to $g$ is given by $\frac{\delta}{\delta g} \int \cc{f} g = \cc{f}$, analogous to the gradient in vector calculus: $\nabla_{\vect{x}} \left(\vect{a}^\dagger \vect{x}\right) = \vect{a}^\dagger$. The primary distinction is the dimensionality of the space and thus the appropriate choice of inner product.

The optimal fields for the various response functions $P$ are therefore those for which $P$ is stationary under small variations of the field degrees of freedom. The field $\vect{F}$ is complex-valued, such that one could take the real and imaginary parts of $\vect{F}$ as independent ($P$ is a nonconstant real-valued functional and therefore not analytic~\cite{Kreutz-Delgado2009} in $\vect{F}$), but a more natural choice is to formally treat the field $\vect{F}$ and its complex-conjugate $\cc{\vect{F}}$ as independent variables~\cite{Remmert1991,Kreutz-Delgado2009}. Then a necessary condition for an extremum of a functional $P\left[\vect{F}\right]$ is for the variational derivatives with respect to the field degrees of freedom to equal zero, $\delta P / \delta \vect{F} = 0$ and $\delta P / \delta \cc{\vect{F}} = 0$ (which are the Euler-Lagrange equations~\cite{Gelfand2000} for functionals that do not depend on the gradients of their arguments). Because our response functions are real-valued, the derivatives with respect to $\vect{F}$ and $\cc{\vect{F}}$ are redundant---they are complex-conjugates of each other~\cite{Kreutz-Delgado2009}---and the condition for the extremum can be found with the single equation
\begin{align}
    \frac{\delta P}{\delta \cc{\vect{F}}} = 0,
    \label{eq:eulerLagrange}
\end{align}
where we have chosen to vary $\cc{\vect{F}}$ instead of $\vect{F}$ for its slightly simpler notation going forward. We apply this variational calculus approach to bound each response function of interest. First, we derive limits for the most general class of materials under consideration. Then we specialize to metals, an important class of lossy media that are typically homogeneous, isotropic, and nonmagnetic at optical frequencies.

\subsection{General lossy media}
We consider first \eqref{pScat}, for the scattered power. Setting the variational derivative of $P_{\rm scat}$ to 0 yields
\begin{align}
    \frac{\delta P_{\rm scat}}{\delta \cc{\vect{F}}} &= -\frac{\varepsilon_0 \omega}{2} \left[ \frac{\tens{\chi}^\dagger \vect{F}_{\rm inc}} {2i} + \left(\Im \tens{\chi}\right) \vect{F} \right] = 0.
    \label{eq:scatDerivGen}
\end{align}
The optimal field that satisfies \eqref{scatDerivGen} is
\begin{align}
    \vect{F}_{\rm scat,opt}(\vect{x}) = \frac{i}{2} \left[\Im \tens{\chi}(\vect{x})\right]^{-1} \tens{\chi}^\dagger(\vect{x}) \vect{F}_{\rm inc}(\vect{x})
    \label{eq:scatOptFieldGen}
\end{align}
for all points $\vect{x}$ within the scatterer volume $V$. The optimal field is guaranteed to exist because $\omega \Im \tens{\chi}$ is positive-definite, per \eqref{lossyDefinition}, and therefore invertible. We have only shown that this is an extremum, not a maximum, but because $\omega \Im \tens{\chi}$ is positive-definite, the scattered power in \eqref{pScat} is a concave functional, for which any extremum must be a global maximum~\cite{Boyd2004}.

One can see that the optimal fields within the scatterer are related to the incident fields (directly proportional for homogeneous media), which conforms intuitively with the scattered-power expression in \eqref{pScat}. The internal field should strongly overlap with the incident field, to increase the power extracted from the incident beam, while the susceptibility dependence balances between maximizing extinction and minimizing absorption. 

A similar procedure yields the optimal internal fields for maximum absorption within a scatterer. Although the absorbed power as given in \eqref{pAbs} is unbounded with respect to $\vect{F}$, adding the constraint that absorption must be smaller than extinction (i.e. the scattered power must be nonnegative) imposes an upper bound. Because \eqref{pAbs} is unbounded, the Karash--Kuhn--Tucker (KKT) conditions~\cite{Nocedal2006} require that the constraint $P_{\rm scat} \geq 0$ must be \emph{active}, i.e. $P_{\rm scat} = 0$. Following standard constrained-optimization theory~\cite{Nocedal2006}, we define the Lagrange multiplier $\ell$ and the Lagrangian functional $\mathcal{L} = P_{\rm abs} + \ell P_{\rm scat}$. The extrema of $\mathcal{L}$ satisfy
\begin{align}
    \frac{\delta \mathcal{L}}{\delta \cc{\vect{F}}} &= \frac{\delta P_{\rm abs}}{\delta \cc{\vect{F}}} + \ell \frac{\delta P_{\rm scat}}{\delta \cc{\vect{F}}} \nonumber \\
    &= \frac{\varepsilon_0 \omega}{2} \left[ (1 - \ell) \left(\Im \tens{\chi}\right) \vect{F} - \frac{\ell}{2i} \tens{\chi}^\dagger \vect{F}_{\rm inc} \right] = 0.
    \label{eq:absDerivGen}
\end{align}
To simultaneously ensure that the scattered power also equals 0, one can verify that the Lagrange multiplier is given by $\ell = 2$. Then the optimal internal fields are
\begin{align}
    \vect{F}_{\rm abs,opt}(\vect{x}) = i \left[\Im \tens{\chi}(\vect{x})\right]^{-1} \tens{\chi}^\dagger(\vect{x}) \vect{F}_{\rm inc}(\vect{x})
    \label{eq:absOptFieldGen}
\end{align}
which are precisely double the optimal scattering fields of \eqref{scatOptFieldGen}. Maximizing $P_{\rm abs}$ subject to $P_{\rm scat} \geq 0$ is a problem of maximizing a convex functional subject to a convex quadratic constraint, such that the solution in \eqref{absOptFieldGen} must be a global maximum~\cite{Urruty2001}.

The limits to the scattered and absorbed powers are given by substituting the optimal fields in \eqref{scatOptFieldGen} and \eqref{absOptFieldGen} into \eqref{pScat} and \eqref{pAbs}, respectively:
\begin{subequations}
\begin{align}
    P_{\rm scat} \leq \frac{\varepsilon_0 \omega}{8} \int_V \cc{\vect{F}}_{\rm inc} \cdot \tens{\chi}^\dagger \left(\Im \tens{\chi}\right)^{-1} \tens{\chi} \vect{F}_{\rm inc} \, {\rm d}^3\vect{x} \label{eq:scatOptGen} \\
    P_{\rm abs}, P_{\rm ext} \leq \frac{\varepsilon_0 \omega}{2} \int_V \cc{\vect{F}}_{\rm inc} \cdot \tens{\chi}^\dagger \left(\Im \tens{\chi}\right)^{-1} \tens{\chi} \vect{F}_{\rm inc} \, {\rm d}^3\vect{x} \label{eq:absOptGen}
\end{align}
\end{subequations}
where extinction has the same limit as absorption, which can be derived by maximing $P_{\rm ext}$ subject to $P_{\rm scat} \geq 0$. The limits depend only on the intensity of the incident field and the material susceptibility $\tens{x}$ over the volume of the scatterer. The product $\tens{\chi} \left(\Im \tens{\chi}\right)^{-1} \tens{\chi}^\dagger$, discussed further below, sets the bound on how large the induced currents can be in a dissipative medium. Whereas optimal per-volume scattering occurs under a condition of equal absorption and scattering, optimal per-volume absorption occurs in the absence of any scattered power and can be larger by a factor of four. This ordering is reversed in the spherical-multipole limits~\cite{Pozar2009,Liberal2014}, where the scattering cross-section (\emph{not} normalized by volume) can be four times larger than the absorption cross-section.

Note that \eqreftwo{scatOptGen}{absOptGen} look superficially similar to the absorbed- and scattered-power limits in \citeasnoun{Hugonin2015}. As discussed in \secref{intro}, they share a common origin as energy-conservation principles applied to integral equations. The key distinction is which quantity serves as a non-negative quadratic (in $\vect{F}$) constraint. We treat absorption as the quadratic quantity, given by \eqref{pAbs}, with the scattered power as the difference between extinction and absorption. The scattered-field-operator approach rewrites the scattered power via a volume integral equation (VIE)~\cite{Chew1995}. This yields a non-negative, quadratic scattered power that is of the same form as \eqref{pAbs} except with the replacement $\Im \tens{\chi} \rightarrow \Im \mathcal{G}$, where $\mathcal{G}$ is a scattered-field integral operator with the homogeneous Green's function as its kernel (the electric component of $\mathcal{G}$ is used in \eqref{scat_int_PM}). Energy conservation leads to the limits of \citeasnoun{Hugonin2015}, which take a similar form to \eqreftwo{scatOptGen}{absOptGen}, except $\Im \tens{\chi} \rightarrow \Im \mathcal{G}$ and the factors of four are reversed. The limits in \citeasnoun{Hugonin2015} are a generalization of the spherical-multipole limits of antenna theory~\cite{Pozar2009,Liberal2014}, which treat the special case of the scattered field decomposed into spherical harmonics.

Our limits have very different characteristics from those of~\cite{Pozar2009,Liberal2014,Hugonin2015}. Our approach, via absorption as the quadratic constraint, yields limits that incorporate the material properties and are independent of structure. This naturally results in per-volume limits, a normalization of inherent interest to designers. The scattered-field-operator approach yields limits that are independent of material but depend on the structure, in a way that can be difficult to be quantified because the inverse of the scattered-field operator is not known except for the simplest cases (e.g. dipoles). A spherical-harmonic decomposition of the operator yields analytical limits, but only to the cross-sections, without normalization. The cross-section is inherently unbounded (increasing linearly with the geometric cross-section at large sizes) and thus difficult to use from a design perspective. The different normalizations are responsible for the different orderings of the absorbed- and scattered-power limits. In \secref{ext_disc} we extend this comparison to show that our material-dissipation approach provides better design criteria at optical frequencies.

The same derivations lead to optimal fields and upper bounds for the radiative and nonradiative LDOS. The optimal fields are nearly identical in form:
\begin{subequations}
\begin{align}
    \vect{F}_{s_j,\textrm{rad,opt}}(\vect{x}) &= \frac{i}{2} \left(\Im \tens{\chi}\right)^{-1} \tens{\chi}^\dagger \cc{\vect{\widetilde{F}}}_{\textrm{inc},s_j}(\vect{x}) \label{eq:rhoRadOptFieldGen} \\
    \vect{F}_{s_j,\textrm{nr,opt}}(\vect{x}) &= i \left(\Im \tens{\chi}\right)^{-1} \tens{\chi}^\dagger \cc{\vect{\widetilde{F}}}_{\textrm{inc},s_j}(\vect{x})
    \label{eq:rhoNrOptFieldGen}
\end{align}
\end{subequations}
where the complex conjugation arises due to the lack of conjugation in the LDOS expressions (which itself arises because open scattering problems in electromagnetism are complex-symmetric rather than Hermitian~\cite{OskooiJo13-sources}). Substituting the optimal fields into the LDOS expressions \eqreftwo{rhoNr}{rhoRad} gives the LDOS limits
\begin{subequations}
\begin{align}
    \frac{\rho_{\rm rad}}{\rho_0} &\leq 1 + \frac{\pi}{2 k^3} \sum_j \int_V \vect{\widetilde{F}}_{\textrm{inc},s_j} \cdot \tens{\chi}^\dagger \left(\Im \tens{\chi}\right)^{-1} \tens{\chi} \cc{\vect{\widetilde{F}}}_{\textrm{inc},s_j} {\rm d}^3\vect{x} \label{eq:rhoRadLimitGen} \\
    \frac{\rho_{\rm nr}}{\rho_0}, \frac{\rho_{\rm tot}}{\rho_0} &\leq \frac{2\pi}{k^3} \sum_j \int_V \vect{\widetilde{F}}_{\textrm{inc},s_j} \cdot \tens{\chi}^\dagger \left(\Im \tens{\chi}\right)^{-1} \tens{\chi} \cc{\vect{\widetilde{F}}}_{\textrm{inc},s_j} {\rm d}^3\vect{x}. \label{eq:rhoNrLimitGen}
\end{align}
\end{subequations}
    %\frac{\rho_{\rm rad}}{\rho_0} &\leq 1 + \frac{\pi}{2 k^3} \sum_j \int_V \vect{\widetilde{F}}_{\textrm{inc},s_j} \cdot \left(\Im \tens{\xi}\right)^{-1} \cc{\vect{\widetilde{F}}}_{\textrm{inc},s_j} {\rm d}^3\vect{x} \label{eq:rhoRadLimitGen} \\
As for extinction, the limit to the total LDOS is identical to the limit to the nonradiative LDOS, which can be proven by maximizing $\rho_{\rm tot}$ subject to $\rho_{\rm rad}\geq 0$.

The absorption, scattering, and LDOS limits in Eqs.~(\ref{eq:scatOptGen},\ref{eq:absOptGen},\ref{eq:rhoRadLimitGen},\ref{eq:rhoNrLimitGen}) depend on the overlap integral of the material susceptibility and the incident field. We can simplify the limits further by separating the dependencies, which is simple for homogeneous, isoptropic media but can also be done for more general media through induced matrix norms~\cite{Trefethen1997}. The integrand in \eqref{scatOptGen} (and each of the other limits) is of the form $\vect{z}^\dagger \bm{A} \vect{z}$, a quantity related to the \emph{norm} (i.e. ``magnitude'') of a matrix $\bm{A}$. The induced 2-norm of a matrix $\bm{A}$, denoted $\|\bm{A}\|_2$, is given by the maximum value of the quantity $\vect{z}^\dagger \bm{A} \vect{z} / \vect{z}^\dagger \vect{z}$ for all $\vect{z} \neq \vect{0}$.  The integral in \eqref{scatOptGen} can then be bounded for general media and arbitrary incident fields:
\begin{align}
    \int_V \cc{\vect{F}}_{\rm inc} \cdot \tens{\chi}^\dagger \left(\Im \tens{\chi}\right)^{-1} \tens{\chi} \vect{F}_{\rm inc} &\leq \int_V \left\| \tens{\chi}^\dagger \left(\Im \tens{\chi}\right)^{-1} \tens{\chi} \right\|_2 \cc{\vect{F}}_{\rm inc} \cdot \vect{F}_{\rm inc} \nonumber \\
     &\leq \left( \left\| \tens{\chi}^\dagger \left(\Im \tens{\chi}\right)^{-1} \tens{\chi} \right\|_2 \right)_{\rm max} \int_V \left|\vect{F}_{\rm inc}\right|^2 \label{eq:xi_bound_norm}
\end{align}
where the dependence on the material susceptibility is now separated from the properties of the incident field $\vect{F}_{\rm inc}$. The field intensity $\left|\vect{F}_{\rm inc}\right|^2$ is proportional to the energy density of the incident field:
\begin{align}
    \frac{1}{2}\varepsilon_0 \left|\vect{F}_{\rm inc}\right|^2 &= \frac{1}{2} \varepsilon_0 \left|\vect{E}_{\rm inc}\right|^2 + \frac{1}{2} \mu_0 \left|\vect{H}_{\rm inc}\right|^2 \nonumber \\
                                                               &= U_{E,\textrm{inc}} + U_{H,\textrm{inc}}
\end{align}
where $U_{E,\textrm{inc}}$ and $U_{H,\textrm{inc}}$ are the (spatially varying) incident electric and magnetic energy densities~\cite{Jackson1999}. Generally the incident fields relevant to $P_{\rm scat}$ and $P_{\rm abs}$ are beams with nearly constant intensity and infinite total energy, for which one should bound the scattered or absorbed power per unit volume of material. Given the operator definition and energy-density relation just discussed, the scattering and absorption limits in \eqreftwo{scatOptGen}{absOptGen} simplify:
%\begin{empheq}[box=\widefbox]{align}
\begin{subequations}
\begin{align}
    \frac{P_{\rm scat}}{V} &\leq \frac{\omega \left(U_{E,\textrm{inc}} + U_{H,\textrm{inc}}\right)_{\rm avg}}{4} \left(\left\|\tens{\chi}^\dagger \left(\Im \tens{\chi}\right)^{-1} \tens{\chi} \right\|_2 \right)_{\rm max} \label{eq:optScatU}  \\
    \frac{P_{\rm abs}}{V}, \frac{P_{\rm ext}}{V} &\leq \omega \left(U_{E,\textrm{inc}} + U_{H,\textrm{inc}}\right)_{\rm avg} \left(\left\|\tens{\chi}^\dagger \left(\Im \tens{\chi}\right)^{-1} \tens{\chi} \right\|_2\right)_{\rm max}. \label{eq:optAbsU}
\end{align}
\end{subequations}
%\end{empheq}

Plane waves are incident fields of general interest. They have equal electric and magnetic energy densities and constant intensities $I_{\rm inc} = c U_{E,\textrm{inc}}$, where $c$ is the speed of light in vacuum. The cross-section of a scatterer is defined $\sigma = P / I_{\rm inc}$, representing the effective area the scatterer presents to the plane wave. Because plane waves are constant in space, the absorption and scattering bounds are tighter. In the first line of \eqref{xi_bound_norm}, $\left|\vect{F}_{\rm inc}\right|^2$ can be taken out of the integral, which then simplifies to the \emph{average} value of the norm of $\tens{\chi}^\dagger \left(\Im \tens{\chi}\right)^{-1} \tens{\chi}$. With this modification to \eqreftwo{optScatU}{optAbsU}, the bounds on absorption and scattering cross-sections per unit volume are:
%\begin{align}
\begin{subequations}
\begin{empheq}[box=\widefbox]{align}
    \frac{\sigma_{\rm scat}}{V} &\leq \frac{k}{2} \left(\left\| \tens{\chi}^\dagger \left(\Im \tens{\chi}\right)^{-1} \tens{\chi} \right\|_2\right)_{\rm avg} \label{eq:optScatSigma} \\
    \frac{\sigma_{\rm abs}}{V}, \frac{\sigma_{\rm ext}}{V} &\leq 2k \left( \left\| \tens{\chi}^\dagger \left(\Im \tens{\chi}\right)^{-1} \tens{\chi} \right\|_2 \right)_{\rm avg}. \label{eq:optAbsSigma}
\end{empheq}
\end{subequations}
%\end{align} 
which apply for general 6$\times$6 electric and magnetic susceptibility tensors. For susceptibilities that are only electric or only magnetic, and therefore 3$\times$3 tensors, the bound is smaller by a factor of two, since the incident magnetic field cannot drive magnetic currents (or vice versa). The LDOS analogue of \eqreftwo{optScatSigma}{optAbsSigma} is not straightfoward, because the incident fields are inhomogeneous. Consequently, we leave \eqreftwo{rhoRadLimitGen}{rhoNrLimitGen} as the general LDOS limits for inhomogeneous media, and derive a simpler version for metals in the next subsection.

Nanoparticle scattering and absorption are often written in terms of electric/magnetic polarizabilities and higher-order moments~\cite{Bohren1983,Hamam2007,Capolino2009,Liberal2014}, whereas \eqreftwo{optScatSigma}{optAbsSigma} are bounds in terms of only the material susceptibility. One implication is that \eqreftwo{optScatSigma}{optAbsSigma} imply restrictions on the number of moments that can be excited, or the strengths of the individual excitations, in a lossy scatterer. A lossy scatterer of finite size cannot have arbitrarily many spherical-multipole moments excited, nor can a single scatterer of very small size achieve full coupling to the lowest-order electric and magnetic dipole moments. Scatterers for which $\Im \chi / \left|\chi\right|^2 \gg V/\lambda^3$ cannot achieve $\sim\lambda^2$ cross-sections per ``channel,'' even on resonance.

\subsection{Metals}
\label{sec:metal_limits}
Metals represent an important and prevalent example of lossy media. (We define a material to behave as a ``metal'' at a given frequency $\omega$ if $\Re \chi(\omega) < -1$, thus including materials such as SiC~\cite{Spitzer1959,Palik1998} and SiO$_2$~\cite{Palik1998} that support surface-phonon polaritons at infrared wavelengths.) At optical frequencies, common metals have homogeneous, isotropic, and nonmagnetic susceptibilities, enabling us to write the matrix norm of the previous subsection as a simple scalar quantity,
\begin{align}
    \left\| \tens{\chi}^\dagger \left(\Im \tens{\chi}\right)^{-1} \tens{\chi} \right\|_{2} = \frac{\left|\chi(\omega)\right|^2}{\Im \chi(\omega)},
\end{align}
where $\chi(\omega)$ is the electric susceptibility. Another alteration in the metal case is that the incident magnetic energy density, $U_{H,\textrm{inc}}$, drops out of the limits because the magnetic polarization currents are zero in \eqref{vol_currents} (and therefore one can simplify $\vect{F}_{\rm inc},\widetilde{\vect{F}}_{\rm inc} \rightarrow \vect{E}_{\rm inc}$). This is not a quasistatic restriction to small objects that only interact with the incident electric field; larger objects that potentially interact strongly with the magnetic field remain valid. But their optimal response can be written in terms of only the incident electric field, since absorption and extinction by nonmagnetic objects can also be written only in terms of electric fields, per \eqreftwo{pAbs}{pExt}. 
 
The limits to per-volume absorption and scattering, simplifying \eqreftwo{optScatU}{optAbsU}, are
%\begin{align}
\begin{subequations}
\begin{empheq}[box=\widefbox]{align}
    \frac{P_{\rm scat}}{V} &\leq \frac{\omega U_{E,\textrm{inc}} }{4} \frac{ \left|\chi(\omega)\right|^2 }{ \Im \chi(\omega) } \label{eq:optScatUMetal}  \\
    \frac{P_{\rm abs}}{V} &\leq \omega U_{E,\textrm{inc}} \frac{ \left|\chi(\omega)\right|^2 }{ \Im \chi(\omega) }. \label{eq:optAbsUMetal}
\end{empheq}
\end{subequations}
%\end{align}
Similarly, the cross-section limits (reduced by a factor of two relative to \eqreftwo{optScatSigma}{optAbsSigma} because $U_{H,\textrm{inc}}$ is not in the limit) are
\begin{subequations}
\begin{empheq}[box=\widefbox]{align}
    \frac{\sigma_{\rm scat}}{V} &\leq \frac{k}{4} \frac{ \left|\chi(\omega)\right|^2 }{ \Im \chi(\omega) } \label{eq:optScatSigmaMetal}  \\
    \frac{\sigma_{\rm abs}}{V}, \frac{\sigma_{\rm ext}}{V} &\leq k \frac{ \left|\chi(\omega)\right|^2 }{ \Im \chi(\omega) }, \label{eq:optAbsSigmaMetal}
\end{empheq}
\end{subequations}
where as before $k = \omega / c$. Whereas the optimal per-volume scattering occurs at a condition of equal absorption and scattering, in \appref{supp_abs} we also derive limits under a constraint of suppressed absorption, as may be desirable e.g. in a solar cell enhanced by plasmonic scattering~\cite{Atwater2010}.

The limits to the power expended by a nearby dipole emitter can be similarly simplified for metals. The incident field  is the field of an electric dipole in free space, proportional to the product of the homogeneous Green's function and the dipole polarization vector, $\vect{F}_{\textrm{inc},s_j} = \vect{E}_{\textrm{inc},s_j} = \tens{G^{EP}} \hat{\vect{s}}_j$ (because the metal is nonmagnetic, only the incident electric field is relevant). The integral over the incident field in \eqreftwo{rhoRadLimitGen}{rhoNrLimitGen}, summed over dipole orientations, is given by $\sum_j \int_V \left|\vect{E}_{\textrm{inc},s_j}\right|^2 = \int_V \left\|\tens{G^{EP}}\right\|_F^2$, where $\|\cdot\|_F$ denotes the Frobenius norm~\cite{Trefethen1997}. For the homogeneous photon Green's function the squared Frobenius norm is shown in \appref{GF_frob_norm} to be 
\begin{align}
    \left\|\tens{G^{EP}}\right\|_F^2 = \frac{k^6}{8\pi^2} \left[ \frac{3}{\left(kr\right)^6} + \frac{1}{\left(kr\right)^4} + \frac{1}{\left(kr\right)^2} \right],
    \label{eq:GEP_frob_norm}
\end{align}
where the $1/r^6$ and $1/r^4$ terms arise from near-field nonradiative evanescent waves, and the $1/r^2$ term corresponds to far-field radiative waves. 

Inserting \eqref{GEP_frob_norm} into the LDOS limits, \eqreftwo{rhoRadLimitGen}{rhoNrLimitGen}, yields a complicated integral that depends on the exact shape of the body. The integrand is positive, though, so one can instead calculate a limit by integrating over a larger space that encloses the body (we will show that most of the potential for enhancement occurs very close to the emitter, such that the exact shape of the enclosure is usually irrelevant). We consider in detail the case in which the scatterer is contained within a half-space, but we also note immediately after \eqreftwo{rhoRadLimitMetal}{rhoNrLimitMetal} the necessary coefficient replacement if the enclosure is a spherical shell. All structures separated from an emitter must fit into a spherical shell, and thus we have not imposed any structural restrictions (in particular, there is no need for a separating plane between the emitter and the scatterer).

We consider a finite-size approximation of the half-space: a circular cylinder enclosing the metal body, a distance $d$ from the emitter and with equal height and radius, $L$ (ultimately we are interested in the limit $L\rightarrow \infty$). The volume integrals are straightforward in cylindrical coordinates, yielding $\int_V 1/r^6 = \pi/6d^3$, $\int_V 1/r^4 = \pi/d$, and $\int_V 1/r^2 = \pi \ln(2) L$, for $L \gg d$ (discarding the contributions $\sim d/L$ for the evanescent-wave terms). Then the limits to radiative and nonradiative LDOS rates are
\begin{subequations}
\begin{align}
    \frac{\rho_{\rm rad}}{\rho_0} &\leq \frac{|\chi(\omega)|^2}{\Im \chi(\omega)} \left[ \frac{1}{32\left(k d\right)^3} + \frac{1}{16 k d} + \BigO{kL} \right] + 1 \label{eq:rhoRadLimitMetal} \\
    \frac{\rho_{\rm nr}}{\rho_0} &\leq \frac{|\chi(\omega)|^2}{\Im \chi(\omega)} \left[ \frac{1}{8\left(k d\right)^3} + \frac{1}{4 k d} + \BigO{kL} \right] \label{eq:rhoNrLimitMetal}
\end{align}
\end{subequations}
where $\BigO{\cdot}$ signifies ``Big-O'' notation~\cite{Cormen2009}. Note that the $\BigO{kL}$ terms, which arise from the far-field excitation, diverge as the size $L$ of the bounding region goes to $\infty$, whereas one would expect the near-field excitation to be most important. The $\BigO{kL}$ divergence as $L\rightarrow \infty$ is unphysical: it represents a polarization current that is proportional to the $1/r$ incident field, according to \eqreftwo{rhoRadOptFieldGen}{rhoNrOptFieldGen}, over the entire half-space, maintaining a constant energy flux \emph{within} a lossy medium. Hence, this $\BigO{kL}$ term, while a correct upper bound, is overly optimistic, and the attainable radiative contribution must be non-diverging in $L$. One could attempt to separately bound the evanescent and radiative excitations. However, $L$ also represents the largest interaction distances over which polarization currents contribute to the LDOS (in \eqref{scat_int_PM}, for example); in \appref{vol_integrals} we show that for reasonable interaction lengths $L$ and near-field separations $d$, the contribution of the $\BigO{kL}$ terms is negligible compared to the $1/d^3$ terms (because the divergence is slow). Thus in the near field, where the possibility for LDOS enhancement is most significant, the limits are dominated by the $1/d^3$ terms:
%\begin{align}
\begin{subequations}
\begin{empheq}[box=\widefbox]{align}
    \frac{\rho_{\rm rad}}{\rho_0} &\leq \frac{1}{32\left(k d\right)^3} \frac{|\chi(\omega)|^2}{\Im \chi(\omega)} \label{eq:rhoRadLimitMetalNF} \\
    \frac{\rho_{\rm nr}}{\rho_0}, \frac{\rho_{\rm tot}}{\rho_0} &\leq \frac{1}{8\left(k d\right)^3} \frac{|\chi(\omega)|^2}{\Im \chi(\omega)}. \label{eq:rhoNrLimitMetalNF}
\end{empheq}
\end{subequations}
%\end{align}
A spherical-shell enclosure of solid angle $\Omega$ yields the same result but with the replacement $1/8 \rightarrow \Omega/4\pi$ in each limit. Again we see the possibility for enhancement proportional to $|\chi|^2 / \Im \chi$. There is the additional possibility of near-field enhancement proportional to $1 / (k d)^3$, which arises from the increased amplitude of the incident field at the metal scatterer.

\begin{figure}[htbp]
\centering
\includegraphics[width=0.7\linewidth]{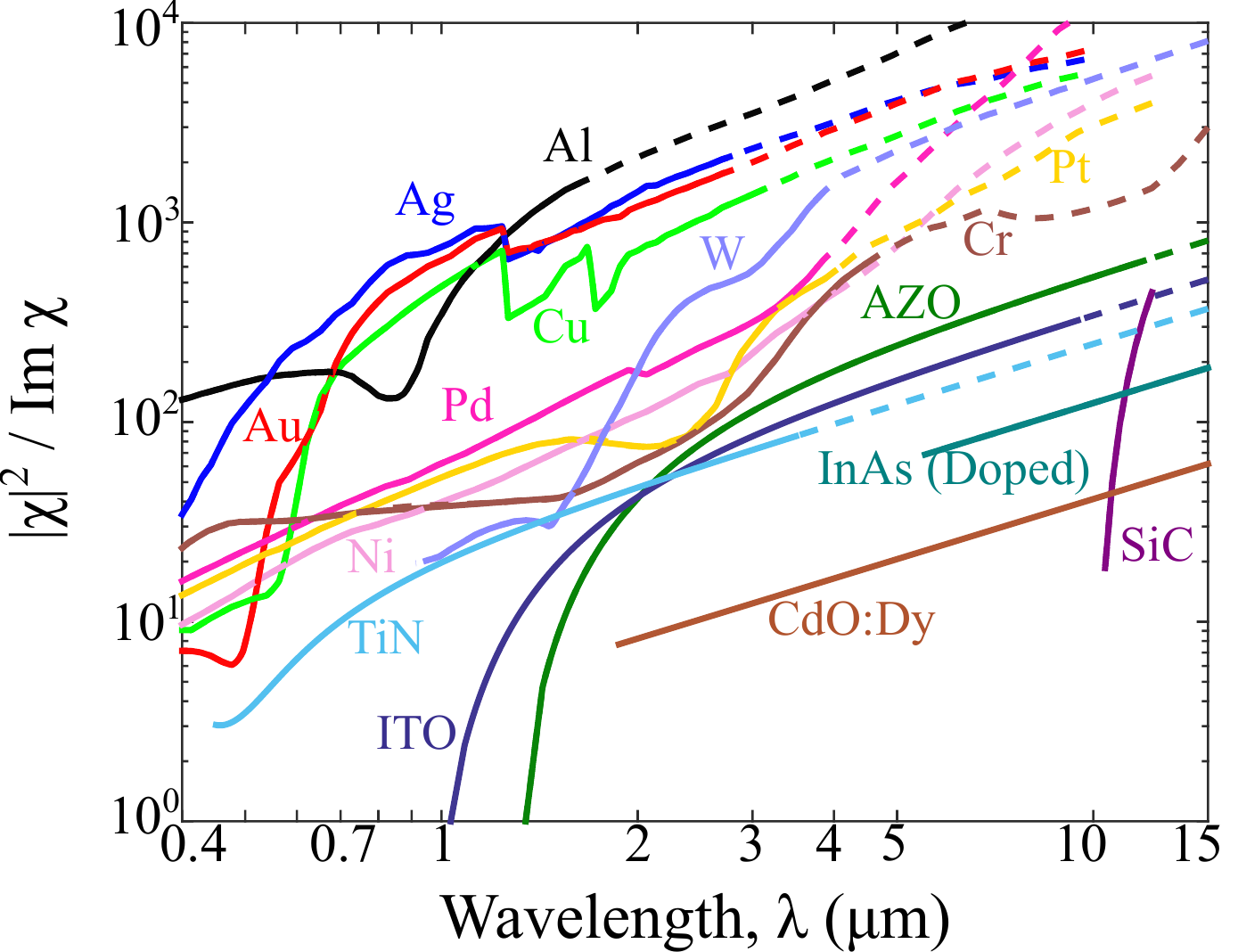}
\caption{\label{fig:materials_by_metric}A comparison of the metric $|\chi|^2 / \Im \chi$, which limits absorption, scattering, and spontaneous emission rate enhancements, for conventional metals (Ag, Al, Au, etc.)~\cite{Palik1998} as well as alternative plasmonic materials including aluminum-doped ZnO (AZO)~\cite{Naik2013}, highly doped InAs~\cite{Law2013}, SiC~\cite{Palik1998}, TiN~\cite{Hibbins1998}, ITO~\cite{Franzen2008}, and Dysprosium-doped cadmium oxide~\cite{Sachet2015} (CdO:Dy). Silver, aluminum, and gold are the best materials at visible and near-infrared wavelengths, although at higher wavelengths the structural aspect ratios needed to achieve the limiting enhancements may not be possible. The dotted lines indicate wavelengths at which resonant nanorods would require aspect ratios greater than 30, approximating the highest feasible experimental aspect ratios~\cite{Busbee2003}. Despite having lower maximum enhancements, AZO, doped InAs, and SiC  should be able to approach optimal enhancements in the infrared with realistic aspect ratios.} 
\end{figure}
\Figref{materials_by_metric} depicts $|\chi|^2 / \Im \chi$ as a function of wavelength for many natural and synthetic~\cite{Palik1998,Hibbins1998,Franzen2008,West2010,Boltasseva2011,Naik2011,Naik2013,Law2013,Sachet2015} metals. Three recent candidates for plasmonic materials in the infrared---aluminum-doped ZnO (AZO), and silicon-doped InAs---are included using Drude models of recent experimental data from Naik et. al.~\cite{Naik2013}, Law et al.~\cite{Law2013}, and Sachet et al.~\cite{Sachet2015}, respectively. For the conventional metals, data from Palik~\cite{Palik1998} was used; high-quality silver, consistent instead with the data from Johnson and Christy~\cite{Johnson1972} and Wu et al.~\cite{Wu2014}, would have smaller losses and a factor of three improvement in $|\chi|^2 / \Im \chi$. A broadband version of the metric can be computed for extinction or LDOS by evaluation at a single complex frequency~\cite{Hashemi2012,Liang2013a}.

The material enhancement factor $|\chi|^2 / \Im \chi$ appears in the absorption cross-section of quasistatic ellipsoids~\cite{Bohren1983}; here we have shown that it more generally bounds the scattering response of a metal of any shape and size. It arises in the increased amplitude of the induced polarization currents; for example, the optimal scattering fields of \eqref{scatOptFieldGen} simplify in metals to the optimal currents
\begin{align}
    \vect{P}_{\rm scat,opt} = \frac{i}{2} \frac{|\chi|^2}{\Im \chi} \varepsilon_0 \vect{E}_{\rm inc},
    \label{eq:optPMetal}
\end{align}
with similar expressions for the optimal currents for maximum absorption and LDOS. The factor $|\chi|^2 / \Im \chi$ provides a balance between absorption and scattering: in terms of the polarization currents, the absorption in a metal is proportional to $(\Im \chi / |\chi|^2) \int_V |\vect{P}|^2$, whereas the extinction is proportional to $\Im \int_V \cc{\vect{E}_{\rm inc}} \cdot \vect{P}$, thus requiring currents proportional to $|\chi|^2 / \Im \chi$ for absorption and extinction to have the same order of magnitude. The expression is intuitively appealing because a large $\left|\chi\right|$ signifies the possibility to drive a large current, while a large $\Im \chi$ dissipates such a current. Our bounds suggest that epsilon-near-zero materials~\cite{Vassant2012,Campione2015}, with $|\chi|\approx 1$, require a very small $\Im \chi$ to generate scattering or absorption as large as can be achieved with more conventional metals.

A similar, alternative understanding can be attained by considering the currents $\vect{J} = d\vect{P} / dt = -i\omega \vect{P}$. Defining the complex resistivity of the metal as $\rho = i / \varepsilon_0 \omega \chi$, the analogue of \eqref{optPMetal} for $\vect{J}$ is
\begin{align}
\vect{J}_{\rm scat,opt} &= \frac{1}{2\Re \rho} \vect{E}_{\rm inc}.
\label{eq:optJMetals}
\end{align}
The enhancement factor $|\chi|^2 / \Im \chi$ thus corresponds to the inverse of the real part of the metal resistivity,
\begin{align}
    \frac{|\chi(\omega)|^2}{\Im \chi(\omega)} = \frac{1}{\varepsilon_0 \omega \Re \rho(\omega)},
\end{align}
which corroborates the idea that small metal resistivities enable large field enhancements, as discussed recently for circuit~\cite{Eggleston2015} and metamaterial~\cite{Tassin2012} models of  single-mode response.

%For single-mode, low-loss structures, the enhancement in radiative LDOS is often described by a  ``Purcell factor'' $Q/V_{\rm eff}$, where $Q$ is the quality factor of the resonance and $V_{\rm eff}$ is the effective mode volume. When the Purcell factor approximation is valid~\cite{OskooiJo13-sources}, the LDOS is related to $Q/V_{\rm eff}$ by~\cite{OskooiJo13-sources,Purcell1946}: $\rho / \rho_0 = 3\lambda^3 Q / 4\pi^2V_{\rm eff}$, and the limits in \eqref{ldosLimit} are thus also limits to $Q/V_{\rm eff}$:
%\begin{align}
    %\frac{Q}{V_{\rm eff}} \leq \frac{\beta}{6\pi} \frac{\left|
    %\chi(\omega) \right|^2}{ \Im \chi(\omega) } \frac{n_0^3}{d^3},
%\label{eq:QVLimit}
%\end{align}
%where the Purcell approximation is only valid for $\rho/\rho_0 \gg 1$ so we have ignored the leading factor of $1$ in \eqref{ldosLimit}.  
The limits in Eqs.~(\ref{eq:optScatSigmaMetal},\ref{eq:optAbsSigmaMetal},\ref{eq:rhoRadLimitMetalNF},\ref{eq:rhoNrLimitMetalNF}) can be applied additively to multiple bodies: the per-volume absorption and scattering limits of \eqreftwo{optScatSigmaMetal}{optAbsSigmaMetal} are equally valid for a single particle, multiple closely spaced particles, layered films, or any other arrangement. Similarly, the LDOS limits in \eqreftwo{rhoRadLimitMetalNF}{rhoNrLimitMetalNF} can be extended to e.g. a structure confined within two half-spaces, but with a prefactor of $2 \times 1/8 = 1/4$, and any other arrangement in space is similarly possible. 

There are two asymptotic limits in which our bounds diverge: lossless metals ($\Im \chi \rightarrow 0$), and, for the LDOS bound, the limit as the emitter--metal separation distance $d \rightarrow 0$. In each case, the divergence is required, as there are structures that exhibit arbitrarily large responses. For example, as the loss rate of a small metal particle goes to zero, it is known that the absorption per unit volume increases until, for a given size, the radiative loss rate equals the absorptive loss rate~\cite{Hamam2007,Ruan2011,Tribelsky2006,Tribelsky2011}.  However, if the size (and therefore the radiation loss rate) is decreased concurrently with the material loss, the cross-section per unit volume can be made arbitrarily large. Thus, for a small enough particle, any $\sigma_{\rm ext}/V$ is possible, and the limit must diverge as material loss approaches zero (regularized physically by both nonzero loss and nonlocal polarization effects~\cite{Scholl2012,Ciraci2012,Eggleston2015}). Similarly, the LDOS can diverge in the limit of zero emitter--scatterer separation, both for lossy materials where absorption diverges~\cite{VanVlack2012} and for lossless materials with sharp corners~\cite{Klimov2001}. The latter case can be reasoned as follows: the fields at a sharp tip, either dielectric or metal, diverge for any nonzero source (of compatible polarization)~\cite{Davis1976,Andersen1978,Jackson1999}. By reciprocity, for a source infinitesimally close to the tip, the LDOS must diverge.

\section{Optimal and non-optimal structures}
\label{sec:structures}
We turn now to the design problem: are there structures that approach the limiting responses set forth by Eqs.~(\ref{eq:optScatSigmaMetal},\ref{eq:optAbsSigmaMetal},\ref{eq:rhoRadLimitMetalNF},\ref{eq:rhoNrLimitMetalNF})? We show that optimal ellipsoids can approach both the absorption and scattering limits across many frequencies by tuning their aspect ratios. For the LDOS limit, however, the optimal designs are not as clear. At the resonant (``surface-plasmon'') frequency $\omega_{\rm sp}$ of a given material the prototypical planar surface exhibits a nonradiative LDOS enhancement approaching the limit of \eqref{rhoNrLimitMetalNF}. However, at lower frequencies, neither thin films~\cite{Miller2014a} nor common metamaterial approaches for tuning the resonant frequency achieve the $|\chi|^2 / \Im \chi$ enhancement, thereby falling short of the limit. Similarly, representative designs for increased radiative LDOS are shown to fall orders of magnitude short of the limits. These structures fall short because the near-field source excites higher-order, non-optimal ``dark'' modes that reduce the LDOS enhancement.

\begin{figure*}
\centering 
\includegraphics[width=\linewidth]{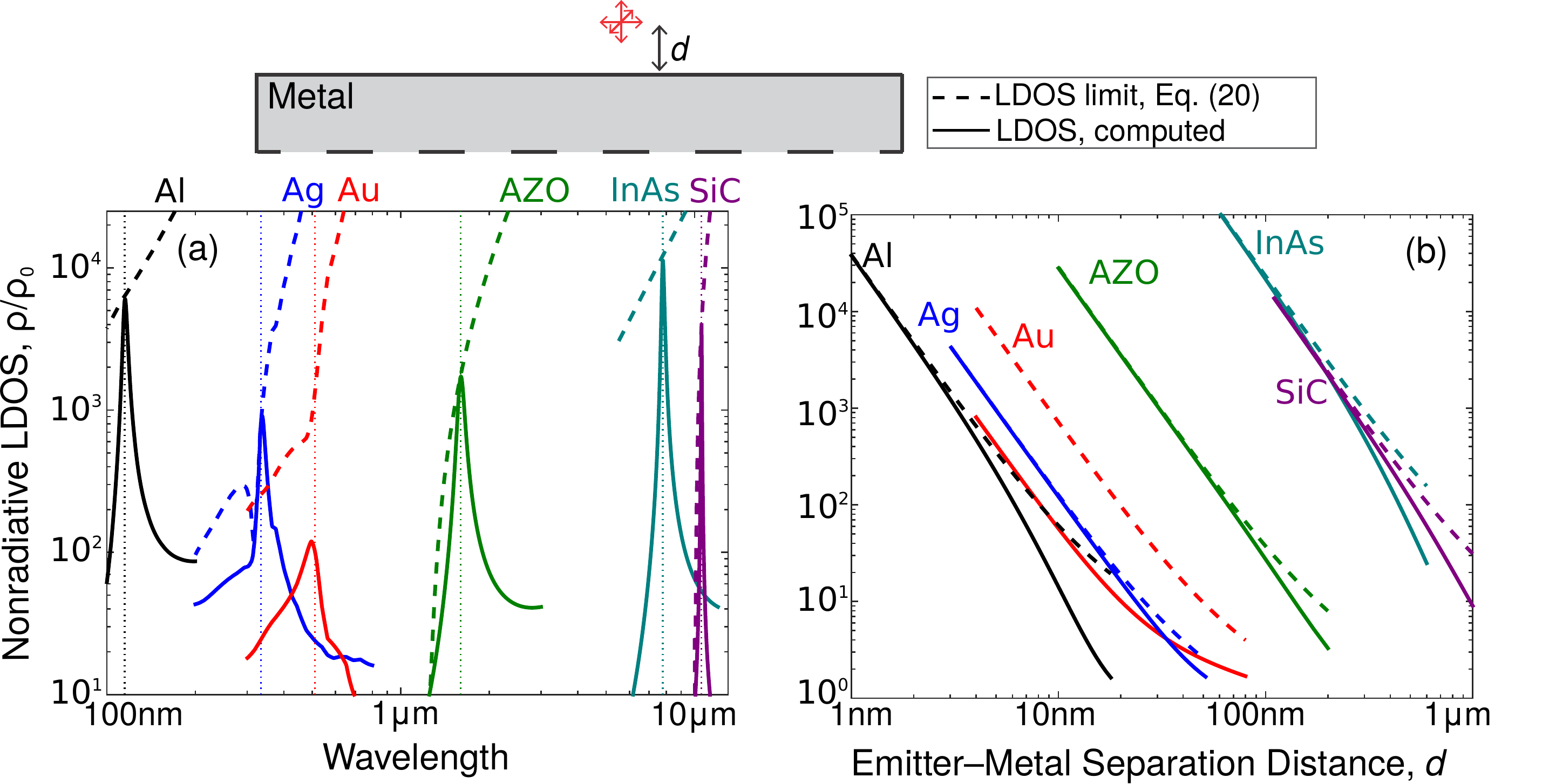}
\caption{\label{fig:ldos_sp_structures} Nonradiative LDOS enhancement for randomly oriented dipoles above a flat bulk metal. (a) Enhancement as a function of wavelength. For each metal except gold---which has significant losses---the nonradiative LDOS at the surface-plasmon frequency $\omega_{\rm sp}$ (dotted line) approaches the limit given by \eqref{rhoNrLimitMetalNF}. The emitter--metal separation distance is fixed at $d = 0.1 c / \omega_{\rm sp}$. The limits are equally attainable for conventional metals such as Al and Ag as for synthetic metals such as AZO~\cite{Naik2013} and highly doped InAs~\cite{Law2013}, and for SiC. (b) Enhancement as a function of metal--emitter separation distance $d$, with the frequency fixed at the surface-plasmon frequency $\omega_{\rm sp}$ for each metal. The limiting enhancements are asymptotically approached as the separation distance is decreased, because the quasistatic approximation of \eqref{ldos_sp} becomes increasingly accurate.
}
\end{figure*}

To compute the electromagnetic response of the structures in Fig.~\ref{fig:extinction_structures}--\ref{fig:ldos_nonsp_structures}, we employed a free-software implementation~\cite{Reid2015,ReidScuffEM} of the boundary element method (BEM)~\cite{Harrington1993}. Where possible (quasistatic ellipsoid extinction~\cite{Bohren1983}, planar metal LDOS~\cite{Sipe1987}), we also used exact analytical and semi-analytical results.

\subsection{Absorption and scattering}
Small ellipsoids approximated by their dipolar response can approach the limits of \secref{limits} across a wide frequency range by tuning their aspect ratios. Ellipsoids reach the absorption limits for small (ideally quasistatic) structures, and the scattering limits for larger structures that are still dominated by their electric dipole moment. The quasistatic absorption cross-section of an ellipsoid, for a plane wave polarized along one of the ellipsoid's axes, is~\cite{Bohren1983}
\begin{align}
    \frac{\sigma_{\rm abs}}{V} = k \Im \left(\frac{\chi(\omega)}{1+L\chi(\omega)}\right),
    \label{eq:sv_ell_qs}
\end{align}
where $L$ is the ``depolarization factor'' (a complicated function of the aspect ratio)~\cite{Bohren1983} along the axis of plane wave polarization. The optimal response is achieved for the aspect ratio such that $L = \Re\left(-1/\chi(\omega)\right)$, which yields a polarization field within the particle of~\cite{Bohren1983}:
\begin{align}
    \vect{P} = i\frac{|\chi|^2}{\Im \chi} \vect{E}_{\rm inc},
\end{align}
which is exactly the optimal absorption condition of \eqref{absOptFieldGen}, as can be seen by comparison with \eqref{optPMetal}. For this optimal depolarization factor, the peak absorption cross-section per unit volume is given by
\begin{align}
    \left[\frac{\sigma_{\rm abs}(\omega)}{V}\right]_{\rm ellipsoid} = k \frac{|\chi(\omega)|^2}
{\Im \chi(\omega)},
    \label{eq:sv_ell_qs_max}
\end{align}
thereby reaching the general limit given by \eqref{optAbsSigmaMetal}. \Eqref{sv_ell_qs_max} is valid for both oblate (disk) and prolate (needle) ellipsoids. Here we have considered the cross-section for a single incident plane wave; if one were interested in averaging the cross-section over all plane-wave angles and polarizations (as appropriate for randomly oriented particles), then it is possible to find a bound that is tighter, by $33\%$ for most materials, in the quasistatic regime. In that case the bounds are achieved by disks but not needles.

Whereas the absorption cross-section is maximized for very small particles approaching the quasistatic limit---necessary to exhibit zero scattered power, a prerequisite for reaching the absorption bounds---the optimal scattering cross-section is achieved for larger, non-quasistatic particles that couple equally to radiation and absorption channels. One can show either through a modified long-wavelength approximation~\cite{Zeman1987,Kelly2003,Moroz2009} or by coupled-mode theory~\cite{Hamam2007,Ruan2011} that the dimensions of a small particle can be tuned such that the absorption and scattering cross-sections are equal, at which point the scattering cross-section per volume is a factor of four smaller than the optimal quasistatic absorption. We validate this result with computational optimization and show that it enables the design of metallic nanorods with nearly optimal performance.

\begin{figure}
    \centering
    \includegraphics[width=\linewidth]{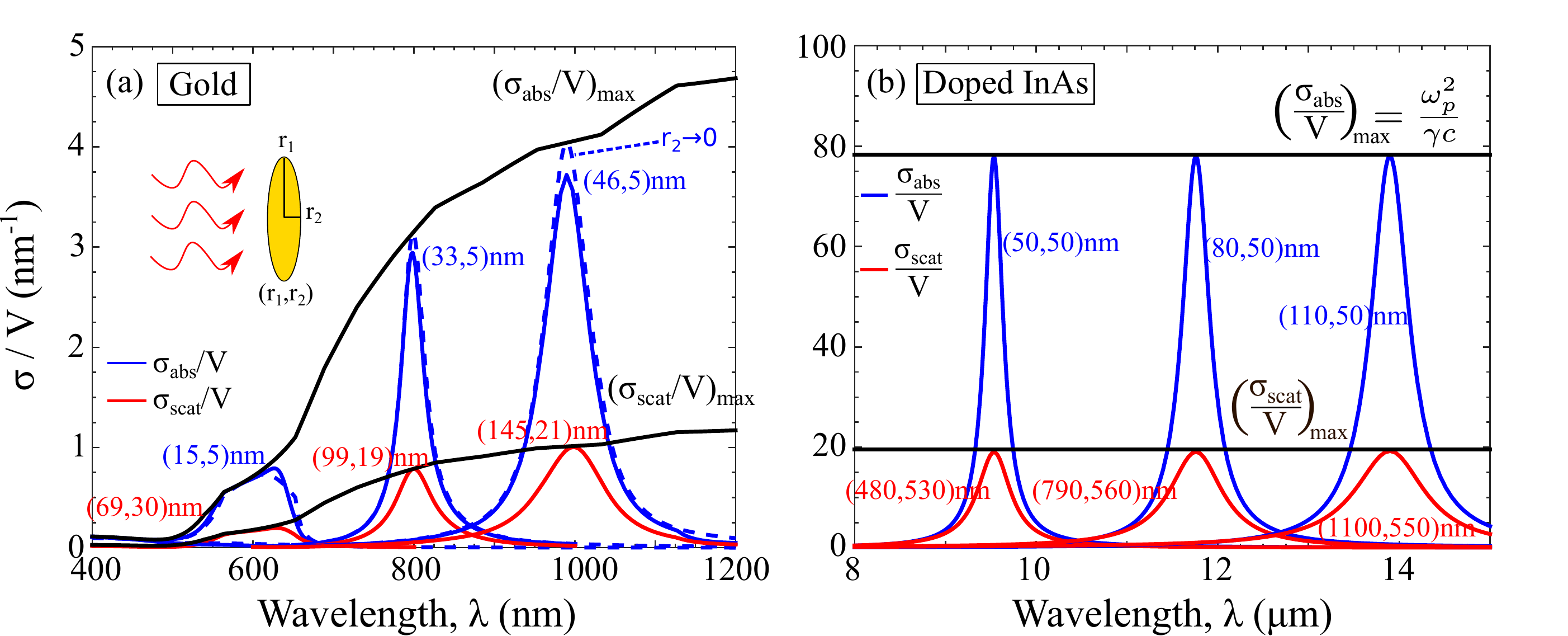}
    \caption{\label{fig:extinction_structures}Absorption (blue) and scattering (red) cross-sections per unit particle volume for nanoparticles of (a) gold~\cite{Palik1998} and (b) Si-doped InAs~\cite{Law2013}, illuminated by plane waves polarized along the particle rotation axis. The ellipsoid aspect ratios can be tuned to approach both the maximum absorption and maximum scattering cross-sections (black) of \eqreftwo{optScatSigmaMetal}{optAbsSigmaMetal}. The dimensions of the nanoparticles are optimized at three representative wavelengths, in the visible for gold (constrained to have radii not less than 5nm) and at longer infrared wavelengths for doped InAs. Whereas the maximum-absorption particles are small to exhibit quasistatic behavior, represented by dashed lines in (a), the maximum-scattering particles are larger such that their scattering and absorption rates are equal.}
\end{figure}
\Figref{extinction_structures} shows the per-volume absorption and scattering cross-sections of (a) gold and (b) Si-doped InAs~\cite{Law2013} nanorods designed for maximum response across tunable frequencies. As in \citeasnoun{Law2013} we employed a Drude model for the doped InAs, with plasma frequency $\omega_p =2\pi c/5.5\mu m$ and damping coefficient $\gamma \approx 0.058\omega_p$, as is appropriate for a doping density on the order of $10^{20}$cm$^{-3}$~(\citeasnoun{Law2013a}). We employed a free-software implementation~\cite{JohnsonNLOpt} of the controlled random search~\cite{Kaelo2006,Price1983} optimization algorithm to find globally optimal ellipsoid radii. For the gold nanorods a minimum radius of 5nm was imposed as representative of experimental feasibility~\cite{Scholl2012,Anquillare2015subm} and a size scale at which nonlocal effects are expected to remain small~\cite{Scholl2012,Ciraci2012,David2014,Eggleston2015}). The gold particles optimized for absorption fall slightly short of the limits due to the minimum-radius constraint; in the quasistatic limit (dashed), the absorption cross-section reaches the limit, as expected from \eqref{sv_ell_qs_max}. Both gold and doped-InAs nanoparticles closely approach the scattering and absoption limits of \eqreftwo{optScatSigmaMetal}{optAbsSigmaMetal}. For Drude models, the factor $k |\chi|^2 / \Im \chi$ that appears in both limits simplifies to $\omega_p^2 / \gamma c$, a material constant independent of wavelength, clearly seen in \figref{extinction_structures}(b). The increase in $\sigma/V$ as a function of wavelength in \figref{extinction_structures}(a), for gold, can be seen as a measure of the deviation of the material response~\cite{Palik1998} from a Drude model. A constant response for Drude models is not universal: \emph{near-field} interactions, in particular the local density of states (LDOS), instead depend only on $|\chi|^2 / \Im \chi \sim \omega_p^2 / \gamma \omega$, thereby increasing at longer wavelengths, away from the bulk and flat-surface plasmon frequencies. 

Not all small particles reach the limiting cross-sections. Coated spheres are common structures for photothermal applications~\cite{Averitt1999,Gobin2007,Dreaden2012}, but their absorption cross-section per unit particle volume is proportional to $(2/3)|\chi|/\Im \chi$ instead of $|\chi|^2 / \Im \chi$ (cf. \appref{coated_sphere_sigma}). Their enhancement does not scale proportional to $|\chi|^2$ due to the small metal volume fractions required to tune the resonant frequency.

\subsection{LDOS}
Designing optimal structures for the local density of states (LDOS) enhancement limits, \eqreftwo{rhoNrLimitMetal}{rhoRadLimitMetal}, is not as straightforward as designing optimal particles for plane-wave absorption or scattering. Because the structure is typically in the near field of the emitter, it is difficult to design a resonant mode that exactly matches the rapidly varying field profile of the emitter. We show that it is possible to reach the nonradiative LDOS limits at the surface-plasmon frequency of a given metal, but that away from these frequencies typical structures fall short. Similarly, for the radiative LDOS limits, common structures fall short of the limits, especially at longer wavelengths.

Planar layered structures support bound surface plasmons that do not couple to radiation, and thus only improve the nonradiative LDOS. As discussed in \secref{intro}, this is potentially useful for radiative heat transfer applications, where near-field emission and absorption have been extensively studied~\cite{Polder1971,Rytov1988,Pendry1999,Ottens2011,Francoeur2008}. Morover, adding either periodic gratings or even random textures can couple the bound modes to the far field~\cite{Nomura2005,Lopez-Tejeira2007,Baron2011} and potentially result in substantial increases to the radiative LDOS. Thus we first study how closely surface modes in planar structures can approach the nonradiative limits, and then we analyze the performance of representative cone- and cylindrical-antenna structures relative to the radiative LDOS limits.

At the surface-plasmon frequency, the prototypical metal-semiconductor interface that supports a surface plasmon exhibits a nonradiative LDOS approaching the limiting value of \eqref{rhoNrLimitMetal}. In the small-separation limit ($kd \ll 1$) and at the surface-plasmon frequency $\omega_{\rm sp}$, the local density of states near a planar metal interface reduces to~\cite{Joulain2003}
\begin{align}
    \left[\frac{\rho_{\rm nr} \left(\omega_{\rm sp}\right)}{\rho_0}\right]_{\rm planar} \approx \frac{ 1 }{ 8\left(kd\right)^3} \frac{|\chi(\omega_{\rm sp})|^2}{\Im \chi(\omega_{\rm sp}) },
\label{eq:ldos_sp}
\end{align}
thereby approaching exactly the nonradiative LDOS limit of \eqref{rhoNrLimitMetal}. (Note that we define $\rho_0$ as the electric-only free-space LDOS, different by a factor of two from the electric+magnetic LDOS in ~\cite{Joulain2003}.) Although the surface-plasmon frequency is typically defined~\cite{Maier2007} as the frequency at which $\Re \varepsilon = -1$, this is the frequency of optimal response only in the zero-loss limit. More generally, we define the surface-plasmon frequency $\omega_{\rm sp}$ such that $\Re \xi(\omega_{\rm sp}) = \Re \left(-1/\chi(\omega_{\rm sp})\right) = 1/2$. For gold, which never satisfies $\Re \xi = 1/2$ due to its high losses, we define surface-plasmon wavelength to be $\lambda_{\rm sp} = 510$nm, where $\Re \xi(\omega)$ is a maximum.
\begin{figure*} 
\centering 
\includegraphics[width=\linewidth]{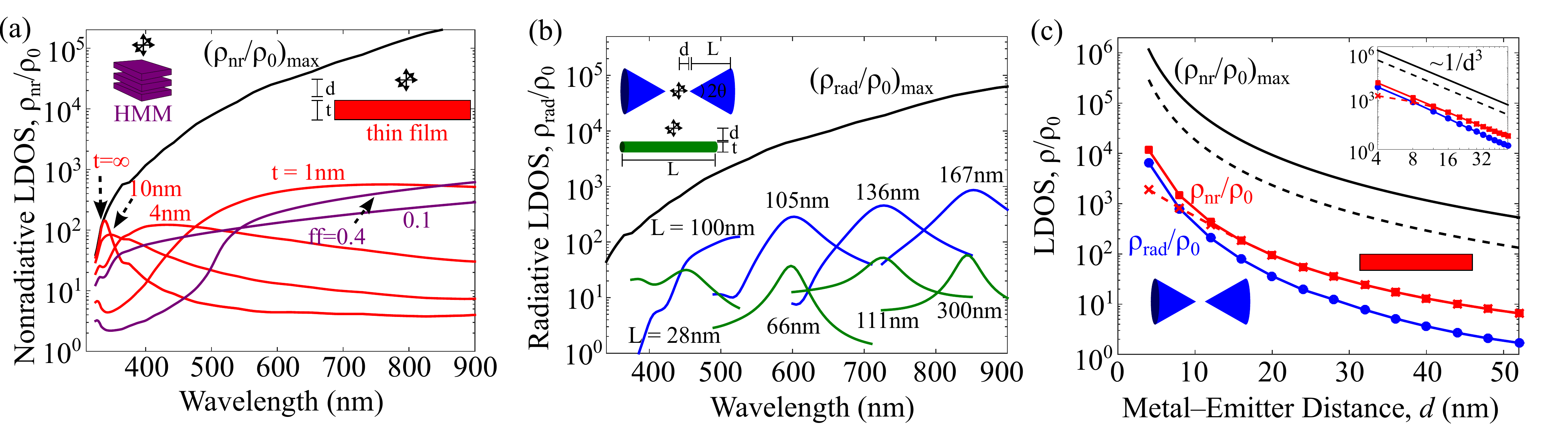}
\caption{\label{fig:ldos_nonsp_structures}Away from the surface-plasmon frequency of a given metal---taken here to be silver~\cite{Palik1998}---it is more difficult to reach the radiative and nonradiative LDOS limits, \eqreftwo{rhoNrLimitMetal}{rhoRadLimitMetal}. The emitter--metal separation $d$ is fixed at $d=10$nm for (a) and (b). (a) Nonradiative LDOS enhancements, $\rho_{\rm nr}/\rho_0$, for thin films (red) for various silver thicknesses ($t$), and for (type-I) hyperbolic metamaterials (HMMs, purple) for two silver fill fractions (ff). (b) Radiative LDOS enhancements, $\rho_{\rm rad}/\rho_0$ for cone and cylinder antennas, with dimensions optimized at wavelengths from $\lambda=450$nm to $\lambda=850$nm. (c) Scaling of $\rho_{\rm nr}$ and $\rho_{\rm rad}$ for optimized thin films and cone antennas, respectively, as a function of $d$ (inset: log--log scale). The scaling of the optimal design appears to be $1/d^3$, with the structures falling short of their respective limits [the dashed line is $(\rho_{\rm rad}/\rho_0)_{\rm max}$] because they do not exhibit a $|\chi|^2 / \Im \chi$ enhancement.}
\end{figure*}

\Figref{ldos_sp_structures}(a) compares semianalytical computations of the nonradiative LDOS near a flat, planar metallic interface to the nonradiative LDOS limits given by \eqref{rhoNrLimitMetal}. Six metals are included: Al (black), Ag (blue), Au (red), AZO (green), InAs (teal), and SiC (purple), with the surface-plasmon frequency $\omega_{\rm sp}$ of each in a dotted line and a fixed emitter--metal spacing of $d = 0.1c/\omega_{\rm sp}$. Every metal except gold---which is too lossy---reaches its respective limit; it is possible that a different, nonplanar gold structure, with the correct ``depolarization factor'' (VIE eigenvalue, cf. \appref{vie_deriv}), could approach the limit. \Figref{ldos_sp_structures}(b) shows the emitter--metal separation-distance dependence for $\omega=\omega_{\rm sp}$. The limits are approached---again, except for gold---as the emitter-metal separation decreases and the approximate $1/d^3$ dependence of \eqref{ldos_sp} becomes more accurate.

There are a few common approaches to tune the resonant frequency below $\omega_{\rm sp}$. A standard approach is to use a thin film~\cite{Economou1969,Maier2007}, coupling the front- and rear-surface plasmons to create lower- and higher-frequency resonances. Other approaches include highly subwavelength structuring, to create hyperbolic~\cite{Krishnamoorthy2012,Cortes2012,Poddubny2013} or elliptical metamaterials with reduced effective susceptibilies. We show here that such structures do {not} exhibit the material enhancement factor, $|\chi|^2 / \Im \chi$, and thus do not approach the limit to $\rho_{\rm nr}$ given by \eqref{rhoNrLimitMetal}.

The nonradiative LDOS of a thin film can be computed by decomposing the dipole excitation into plane waves (including evanescent waves), which reflect from the layers according to the usual Fresnel coefficients. The LDOS near a thin film is well-known as an integral over the surface-parallel wavevector~\cite{Joulain2005}; for a dipole with fixed frequency $\omega$ and height $d$ above the film, and a film of optimal thickness $t$, one can show that $\rho_{\rm nr}$ is (cf. \appref{thin_film_ldos})
\begin{align}
    \left[\frac{\rho_{\rm nr}(\omega)}{ \rho_0 }\right]_{\textrm{thin film}} \approx \frac{1}{2\left(k d\right)^3},
    \label{eq:ldos_thin_film}
\end{align}
which is valid for $\omega << \omega_{\rm sp}$ (otherwise a bulk is optimal and \eqref{ldos_sp} describes the response) and for relatively small loss, $\Im \chi \ll \left|\Re \chi\right|$ (as is typical at optical frequencies), to ensure the large-wavevector modes are resolved. Unlike a planar interface, an optimal thin film does not exhibit the $|\chi|^2 / \Im \chi$ material enhancement. The thin film falls short because it relies on near-field interference to couple the front- and rear-surface plasmons, yielding a resonance that couples strongly to the dipole emitter over only a small bandwidth of wavevectors ($\Delta k_p \sim \Im \chi / |\chi|$) that cancels the resonant enhancement from decreased loss. An alternative understanding arises from viewing a thin film as the single-unit-cell limit of a layered hyperbolic metamaterial (HMM)~\cite{Miller2014a}.  Hyperbolic metamaterials exhibit anisotropic effective susceptibilities such that the resonant frequency can be tuned, but their resonances occur within the bulk rather than along the surface, such that they cannot yield infinite LDOS even in the limit of zero loss. In \citeasnoun{Miller2014a} we show that the LDOS near an optimal HMM is nearly identical to the LDOS near an optimal thin film, as verified in \figref{ldos_nonsp_structures}.

Achieving the radiative LDOS limit is a similarly challenging problem. To reach the radiative LDOS limits, the polarization field must exhibit the $1/r^3$ spatial dependence of the incident field, while also coupling to far-field radiation channels. Here we consider two representative structures for tuning the LDOS resonant frequency: mirror-image cones (akin to bowtie antennas) and cylindrical antennas (scaled~\cite{Novotny2007} shorter than $\lambda/2$). and we show that each falls short of the limits across optical frequencies. 

\Figref{ldos_nonsp_structures} shows the nonradiative and radiative LDOS near silver thin films, effective-medium hyperbolic metamaterials (HMMs), cones, and cylinders. In \figref{ldos_nonsp_structures}(a) and \figref{ldos_nonsp_structures}(b), the emitter--metal separation distance is fixed at $d=10$nm, and the structures are optimized at four wavelengths, $\lambda = [450,600,725,850]$nm, using a standard local optimization algorithm~\cite{Powell1994}. The optimal lengths and thicknesses are included in the figure. The optimal cone half-angles and cylinder radii vary slightly but are typically on the order of $\theta = 20$\textdegree and $r=10$nm, resp. Away from $\omega_{\rm sp}$, the optimal nonradiative LDOS of the thin films in \figref{ldos_nonsp_structures}(a) is very well described by \eqref{ldos_thin_film}, which shows that the structures fall short of the limits by the material enhancement factor. A similar effect is seen for the radiative LDOS in \figref{ldos_nonsp_structures}(b), as the cones also fall increasingly short of the limits as the frequency is decreased ($\approx 250\times$ at $\lambda=850$nm). The quantum efficiencies ($\rho_{\rm rad}/\rho_{\rm tot}$) of the cone antenna varies from 50--70$\%$, near the optimal ratio of $50\%$ to reach our bounds, suggesting that the reason they fall short is due to a mismatch between the emitter and resonance field profiles, not due to coupling to nonradiative channels. \Figref{ldos_nonsp_structures}(c) shows the LDOS dependencies as a function of $d$ for $\lambda=600$nm, in linear (main) and logarithmic (inset) scale, with the structure parameters optimized for each $d$. The optimal structures appear to exhibit the $1/d^3$ scaling of the limits in \eqreftwo{rhoRadLimitMetal}{rhoNrLimitMetal}. An important open question is the extent to which structures can be designed to approach the limits, thereby improving over current designs by two to three orders of magnitude. 

\section{Extensions and discussion}
\label{sec:ext_disc}
\begin{figure*}
\centering
\includegraphics[width=\linewidth]{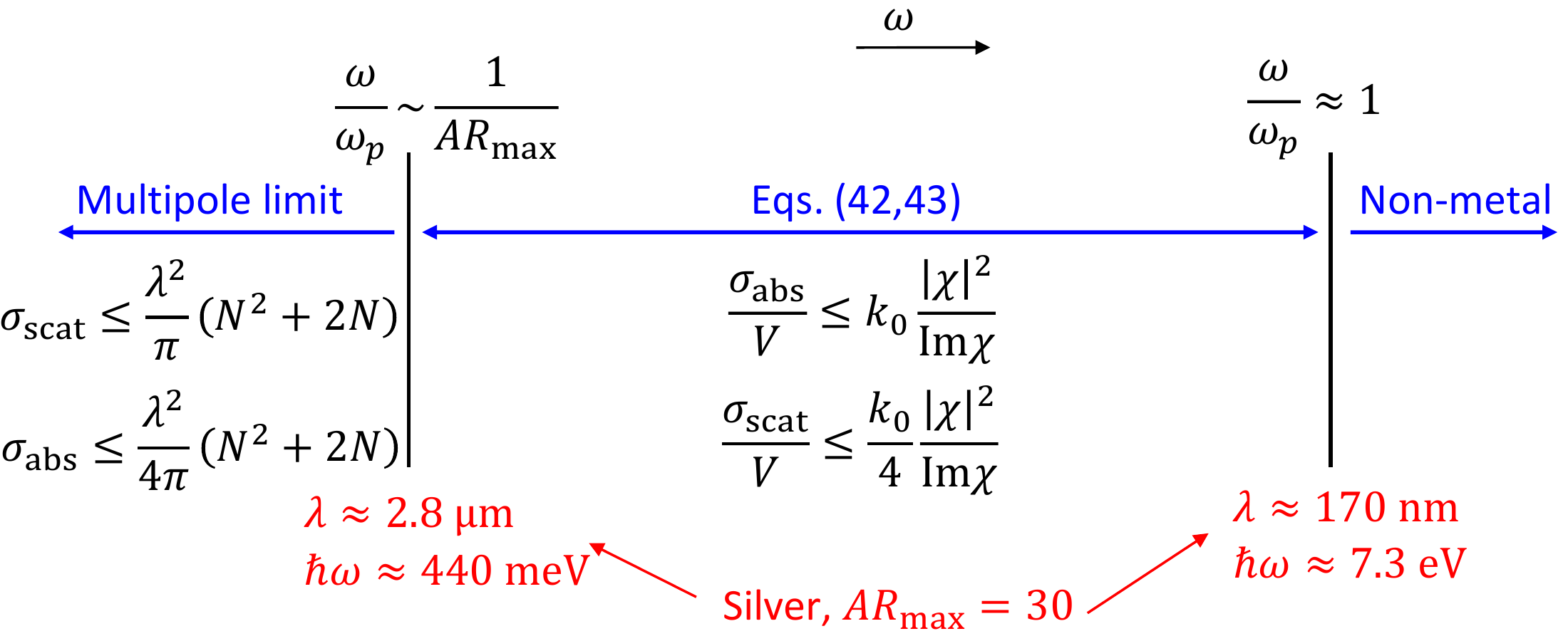}
\caption{\label{fig:limits_comparison}A schematic comparison of absorption and scattering limits: multipole limits~\cite{Pozar2009,Liberal2014} to the total cross-section can provide design guidelines at low frequencies, where it is difficult to achieve ``plasmonic'' resonances, but at higher frequencies our dissipation-based limits provide tighter limits to the per-volume response. The frequencies at which our bounds can be reached range from the bulk plasma frequency ($\omega = \omega_p$) down to $\omega \sim \omega_p/AR_{\rm max}$, where $AR_{\rm max}$ is the maximum achievable aspect ratio. Included is the relevant range for silver ellipsoids (red text), assuming $AR_{\rm max} = 30$. Plasmonic behavior at longer wavelengths is possible with materials such as AZO and doped InAs (cf. \figref{materials_by_metric}).} 
\end{figure*}
The limits derived in Secs.~(\ref{sec:AbsScatLDOS},\ref{sec:limits}) apply to a general class of materials embedded in vacuum, without any scatterers. They can be generalized for non-vacuum backgrounds. For a metal in a homogeneous, lossless background with permittivity $\varepsilon_{\rm bg}$, the limits are of the same form but with replacements $k \rightarrow \sqrt{\varepsilon_{\rm bg}} k$ and $\chi \rightarrow (\varepsilon(\omega)-\varepsilon_{\rm bg})/\varepsilon_{\rm bg}$. A similar generalization applies for general lossy media in non-vacuum backgrounds. If there are background scatterers present (possibly periodic~\cite{Silveirinha2005a,Capolino2007,Shore2007}), then the derivation is identical except that the Green's functions are the Green's functions in the presence of the background scatterers, and the ``incident'' fields therefore incorporate the effects of the background scatterers. The minimum thickness of a metal absorber on a substrate~\cite{Liu2010,Tittl2011}, for example, could be computed with \eqref{optAbsSigmaMetal}, with the replacement $E_{\rm inc} = E_0 \left(e^{ikz} + re^{-ikz} \right)$ (for substrate reflectivity $r$), or more generally by bounding the incident field via $|E_{\rm inc}| < 2 |E_0|$.

Previous approaches to general limits via energy conservation have bounded the response of a structure via its scattered-field operator~\cite{Hugonin2015}. This has yielded limits to the cross-section, $\sigma$, for spherically symmetric~\cite{Hamam2007,Ruan2011} or more general~\cite{Pozar2009,Liberal2014} scatterers whose response has been decomposed into spherical multipoles. The cross-section limits are proportional to $\lambda^2\left(N^2+N\right)$, where $N$ is the number of excited multipoles.

From a design perspective, there are serious impediments to using such cross-section limits (as opposed to our $\sigma/V$ limits). The cross-section itself is unbounded, increasing with the size of a large particle~\cite{Bohren1983}. Thus implicit in any use of such bounds is a size normalization, but this can only be done at very small scales, where the number of multipoles is 1 (or 2 for perfect conductors) and our bounds are tighter, and at very large size scales, approaching the geometric-optics regime. At intermediate sizes, it is difficult to estimate the number of multipoles without further modeling of a given structure. We have presented a new approach, using material dissipation, instead of the scattering operator, as the binding constraint. Our limits have the unique feature that they incorporate the material properties and are independent of structure. From a design viewpoint this is a significant advantage, since for any problem there are infinitely many possible structures but typically only a handful of relevent materials. Furthermore, the normalization to geometric volume, e.g. $\sigma/V$, emerges naturally in our limits.

Reaching our absorption and scattering limits likely requires significant polarization currents throughout the scatterer volume, as can be seen in the optimal field profiles in \secref{limits}. At low frequencies, it is difficult to fabricate structures with sizes or aspect ratios necessary to achieve such resonances. For a Drude-metal (appropriate at low frequencies) ellipsoidal nanorod, the optimal aspect ratio~\cite{Bohren1983} for maximum absorption scales as $\omega/\omega_p$, where $\omega_p$ is the bulk plasma frequency. Thus, if we define a maximum feasible aspect ratio $AR_{\rm max}$, the minimum frequency at which a plasmonic resonance can be achieved is proportional to $\omega_p / AR_{\rm max}$. Below this frequency, the multipole limits can serve as a design guide, although the uncertainty about the number of multipoles, and the potential mismatch between a non-spherical object and its bounding sphere, remain as barriers. \Figref{limits_comparison} depicts schematically depicts which bound provides better design criteria as a function of frequency. Included in \figref{limits_comparison} is red text corresponding to the limiting frequencies at which silver nanorods (using experimental material data~\cite{Palik1998} and making no Drude approximation) can approach our limits, assuming a realistic~\cite{Busbee2003} maximum aspect ratio of 30.  

Similarly, it may be difficult to reach our limits with larger, wavelength-scale solid particles that are much larger than the skin depth. One of the conclusions from our work is that such particles are particularly inefficient scatterers, and thus should be \emph{avoided}, because currents cannot be excited throughout such a large portion of their volume. At optical frequencies, any technology must ultimately incorporate some \emph{collection} of ordered or disordered scatterers, whether in planar arrays~\cite{Nie1997,Landy2008,Pala2009,Atwater2010,Aydin2011,Hsu2014a}, aqueous environments~\cite{Peer2007,Boisselier2009,Qiu2012}, or some other configuration. Thus even if the individual scatterers have small cross-sections, there can be many of them (due to their small volumes), providing a large collective cross-section~\cite{Anquillare2015subm} while maintaining the per-volume response of the individual scatterers.

Aside from particle scattering, our limits extend to situations that do not have multipole counterparts. They yield meaningful limits for extended structures (whose large size would excite many spherical harmonics), and for the LDOS (where a near-field source would excite many spherical harmonics).

An interesting aspect of our limits is the fact that they apply to any open scattering problem. Open systems---in which energy can enter and exit---are typically described by non-Hermitian operators. Non-Hermitian operators are not guaranteed to be diagonalizable, and therefore may not have a complete basis of eigenfunctions (technically, even for Hermitian operators, rigorous eigendecomposition in infinite-dimensional spaces is subtle and subject to obscure counter-examples~\cite{Reed1980,Pedersen1989}). Breakdown of diagonalizability only occurs at ``exceptional points'' that must be forced~\cite{Kato1980,Heiss2004} and which occur by chance with zero probability. Near an exceptional point, however, it is possible to have eigenfunctions that are nearly ``self-orthogonal,'' with exceptionally large modal overlap (or in theory highly nonnormal and ill-conditioned Maxwell operators~\cite{Trefethen2005}) leading to effects such as destructive interference in scattering ``dark states''~\cite{Hsu2014} (e.g. Fano resonances~\cite{Fano1961}) and the Petermann factor for noise enhancement in lasers~\cite{Petermann1979,Siegman1989}. Nevertheless, passive systems near such exceptional points cannot exceed our limits, which impose only conservation of energy.

The limits presented here suggest new design opportunities with metals. Nearly lossless metals~\cite{Khurgin2010} could manifest unprecedented responses. Even for conventional lossy metals, large-area structures that achieved the absorption or scattering limits presented here could potentially do so with thicknesses approaching a single atomic layer. Nonlocal interactions, for which it may not be possible to separate material and structural properties, would be important in such a structure. Finding limits incorporating nonlocal effects would represent an important extension to this work. Designing structures to approach the LDOS limits could impact applications such as imaging, where there are potentially orders of magnitude improvement to be gained. We derive limits for the problem of near-field radiative heat transfer, where the sources are embedded \emph{within} the designable media, in an upcoming publication~\cite{Miller2015}, and it would be interesting to extend the limits derived here to other figures of merit, potentially finding new metrics or structures for optimal light--matter interactions.

\appendix
\section{Alternative understanding of the limits: VIE approach}
\label{app:vie_deriv}
Here we present an alternative viewpoint for understanding our limits, as arising from sum rules over eigenmodes of the volume integral equations of electromagnetism. This approach appears to only work for materials with a scalar $\chi$ (either electric or magnetic), and thus is less general than the derivation in the text (we assume a nonmagnetic medium). We include this appendix because higher-order sum rules may yield tighter limits in certain scenarios, e.g. angle-averaged incident fields. First we show how limits arise from eigenmodes of the volume integral equations (VIEs), which can be considered ``material resonances.'' This connection was partially recognized by Rahola~\cite{Rahola2000}, and may be related to eigenvalues in SALT laser theory~\cite{Esterhazy2014}, but has not since been pursued any further. Material resonances are common in quasistatic electromagnetism~\cite{Ouyang1989,Mayergoyz2005}, where there is no frequency; here, we show how to extend the concept to fixed, nonzero frequencies.

\begin{figure} 
    \centering
    \includegraphics[width=0.7\linewidth]{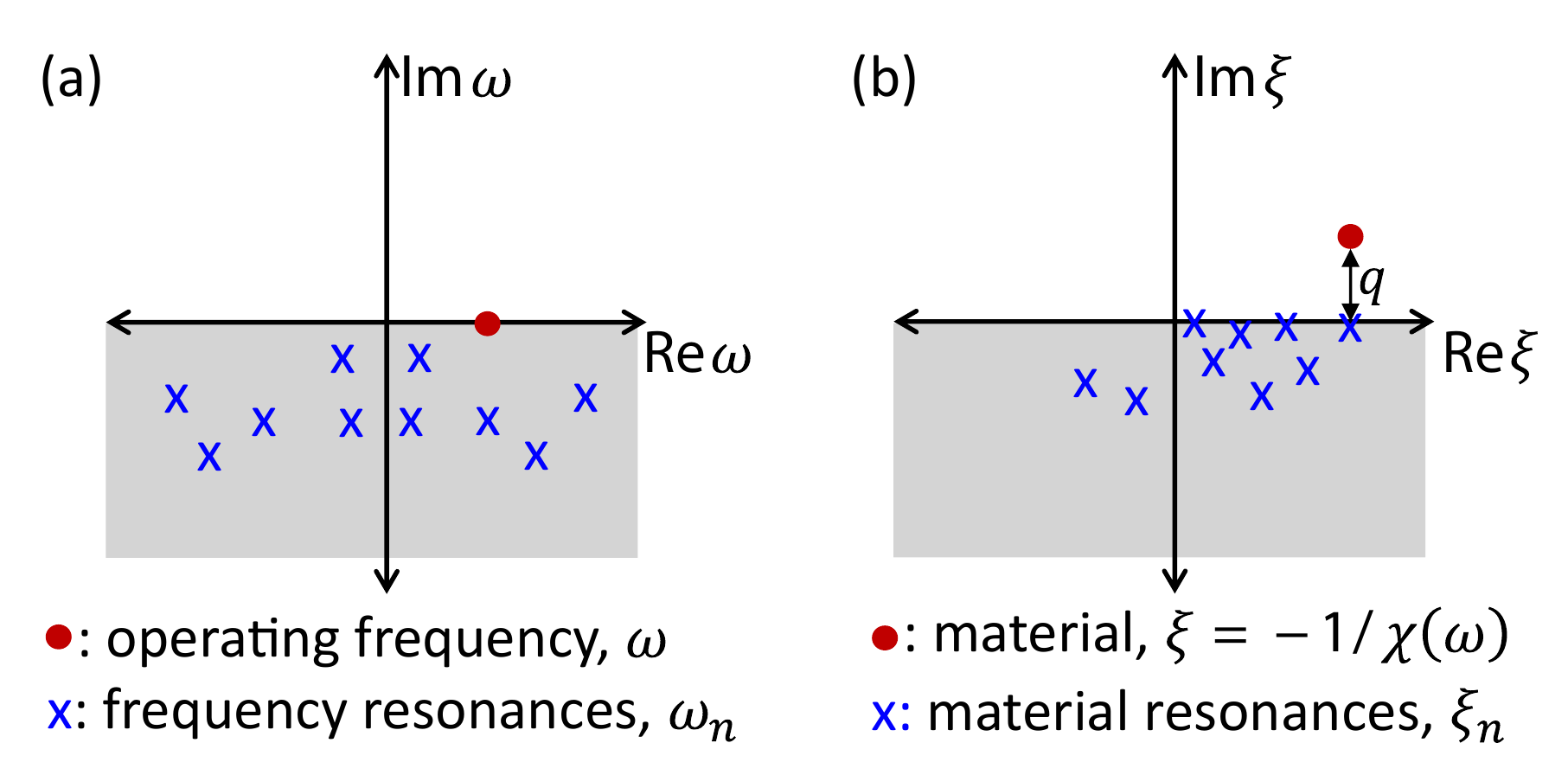}
    \caption{\label{fig:resonance_comp}A comparison of the (a) resonant-frequency and (b) resonant-susceptibility frameworks in electromagnetism. The conventional resonant-frequency approach is depicted in (a): the operating frequency is real-valued and can in theory be approached arbitrarily closely by a resonance with small imaginary part $-\Im \omega_n$ (i.e. a high-Q resonance).  Conversely, volume integral equations yield the resonant-susceptibility approach depicted in (b): metal losses, which correspond to $\Im \xi = \Im \left(-1/\chi\right) > 0$, inherently impose a minimum separation $q$ to how closely a material resonance---restricted to lie on or below the real line---can approach the real system parameters. Moreover, quasistatic structures have real-valued eigenvalues, and thus have the potential to achieve the minimum eigenvalue separation and maximum optical response.}
\end{figure}

Inherent to the concept of a resonance in physics is the resonant frequency: intuitively, the frequency at which an electromagnetic, elastic, quantum mechanical, or any other type of wave oscillates without external forcing in a specific, predefined structure. In a closed or periodic structure, these resonances correspond mathematically to eigenvalues of the underlying differential equations. For photonic structures, defined by a spatially dependent permittivity $\varepsilon(\vect{x})$, resonant frequencies $\omega_n$ are eigenvalues of the eigenequation 
\begin{align}
    \frac{1}{\varepsilon(\vect{x})} \nabla \times \nabla \times \vect{E}_n = \left(\frac{\omega_n}{c}\right)^2 \vect{E}_n,
    \label{eq:Maxwell_Eig}
\end{align}
defined with appropriate boundary conditions and the divergence equation $\nabla \cdot \varepsilon \vect{E} = 0$. For a material with negligible dispersion (metals introduce further complications), the Maxwell operator $\mathcal{M} = \varepsilon^{-1} \nabla \times \nabla \times$ is frequency-independent, such that the eigenvalue corresponds to a resonant frequency. In open systems, resonances are complex poles of the Green's function (or of the scattering operator) rather than true eigenvalues, because their ``eigenfunctions'' (``leaky modes'') diverge exponentially in space~\cite{Lax1990,Zworski1999}. For a sign convention of $e^{-i\omega t}$ time dependence, these poles $\omega_n$ must lie in the lower-half of the complex-frequency plane as shown in \figref{resonance_comp}(a). Frequency resonances in electromagnetism are well understood, but \emph{material} resonances---which arise as eigenvalues in integral equations---have hardly been explored at all.  

The electric field integral equation (EFIE) formulation of Maxwell's equations is derived through the use of Green's functions~\cite{Chew1995,Polimeridis2014}. As depicted in \figref{scattering_problem}, we consider generic scattering problems in which a structure with susceptibility $\chi$ interacts with an externally imposed incident field $\vect{E}_{\rm inc}$. The response of the scatterer is given by a convolution of the free-space Green's function $\tens{G}$ with the induced polarization currents $\vect{P} = \chi \vect{E}$. As in the text we assume a scatterer embedded in vacuum, with straightforward generalizations. The total field $\vect{E}$ is the sum of the incident and scattered fields~\cite{Chew1995,Sun2009},
\begin{align} 
    \vect{E}(\vect{x}) = \vect{E}_{\rm inc}(\vect{x}) - \int_V
    \chi(\vect{x}') \tens{G}(\vect{x},\vect{x}',\omega)\vect{E}(\vect{x}')\,{\rm
    d}V
    \label{eq:VIE0}
\end{align} 
for all points in space, where we choose a negative sign convention for the Green's function (opposite that of \eqref{scat_int_PM} in the text). \Eqref{VIE0} can be desingularized~\cite{Samokhin2001}, but our treatment depends only on an abstract spectral decomposition that does not require us to grapple with such details. A similar integral equation arises in quantum mechanics, where it is known as the Lippmann--Schwinger equation and the susceptibility is replaced by the scattering potential~\cite{Sakurai1994}. For a scatterer with homogeneous susceptibility, $\chi$ is constant and can be taken out of the integrand in \eqref{VIE0}.

Homogeneous scatterers are thus defined by the single material parameter $\chi$. Resonances of \eqref{VIE0}, even in open systems, are true eigenvalues because the integral equation has unknowns $\vect{E}(\vect{x}')$ defined over the finite scatterer domain $V$, and thus the corresponding eigenfunctions are normalizable. Eigenfunctions $\vect{E}_n$ and eigenvalues $\xi_n$ of the Green's function integral operator satisfy:
\begin{align} 
    \int_V \tens{G}(\vect{x},\vect{x}',\omega)
    \vect{E}_n(\vect{x}') \,{\rm d}V = \xi_n \vect{E}_n(\vect{x}) = -\frac{1}{\chi_n}
    \vect{E}_n(\vect{x})
    \label{eq:VIE_Eig}
\end{align} 
for all points $\vect{x}$ in $V$. Given the eigenvalue $\xi_n$, if we choose a material $\chi = \chi_n = -1/\xi_n$ then by comparison with \eqref{VIE0} we see that a ``standing-wave'' $\vect{E} \neq 0$ is possible even for $\vect{E}_{\rm inc} = 0$. \Eqref{VIE_Eig} is the integral-equation analogue of \eqref{Maxwell_Eig} (for a homogeneous scatterer), and yet we see that the integral operator on the left-hand side of \eqref{VIE_Eig} depends on the structure and the frequency but not on the susceptibility. Instead, the eigenvalue of the mode yields a resonant value $\chi_n$---a resonant material susceptibility, for a fixed frequency. Just as ``leaky modes'' in \eqref{Maxwell_Eig} are not actually physical solutions of Maxwell's equations, the resonances $\chi_n$ are not physically realizable materials; we will see below that $\Im \chi_n < 0$ for $\omega>0$, as shown in \figref{resonance_comp}(b), corresponding to gain required to overcome modal radiation loss.

For both frequency and material resonances, the separation between the system parameter (e.g.  operational frequency) and the resonance defines the magnitude of the response---the smaller the separation, the larger the response. In the resonant-frequency framework it is difficult to provide a lower bound on the imaginary part of the resonant frequency $\omega_n$ (thereby bounding the maximum $Q$). In the resonant-material approach, however, the system parameter---the susceptibility, instead of the frequency---has a nonzero imaginary part for a lossy system. By causality~\cite{Landau1960} (or passivity~\cite{Welters2014}), frequency resonances for a fixed structure and material lie in the lower half of the complex-$\omega$ plane. Similarly, for a fixed frequency $\omega>0$, the material resonances $\xi_n$ must reside in the lower half of the complex-$\xi$ plane (otherwise, one could construct a passive linear material that violates the condition on the resonant frequencies~\cite{WeltersPC}). Thus, as depicted in \figref{resonance_comp}, there is a minimum separation $q = \Im \xi(\omega)$ between the material parameter $\xi(\omega) = -1/\chi(\omega)$ and the eigenvalue $\xi_n$. There are further benefits to the resonant-material approach: the solutions to the integral equation are defined only over the scatterers, rather than all space, and quantities like the extinguished power and the local density of states can be written as volume integrals, ideally suited to a VIE framework.  

To simplify further analysis, we rewrite the VIE of \eqref{VIE0} as:
\begin{align}
    \left(\mathcal{I} + \chi \mathcal{G}\right) \vect{e} = \vect{e}_{\rm inc}
    \label{eq:VIE}
\end{align}
where the fields are $\vect{e}$ and $\vect{e}_{\rm inc}$ (with lower-case $\vect{e}$ denoting vector fields restricted to the volume $V$, forming a Hilbert space, as opposed to the fields $\vect{E}$ defined everywhere in space), $\mathcal{I}$ is the identity operator, and $\mathcal{G}$ is the Green's function integral operator defined by $\mathcal{G} \vect{e} = \int_V \tens{G}(\vect{x},\vect{x}',\omega) \vect{E}(\vect{x}')$. The eigenfunctions of $\mathcal{G}$ are solutions of \eqref{VIE_Eig} and are given in vector notation by
\begin{align} 
    \mathcal{G} \vect{e}_n = \xi_n \vect{e}_n,
\end{align}
where $n$ is the mode index, $\xi_n = -1/\chi_n$, and $\Im \xi_n \leq 0$. There are advantages to considering the integral operator $\mathcal{G}$ rather than the Maxwell operator $\mathcal{M} = (1/\varepsilon)\nabla \times \nabla \times$.  For the operator $\mathcal{M}$, metals are difficult to treat: material dispersion renders the eigenvalue problem nonlinear in $\omega^2$, and material loss yields a non-Hermitian operator even for closed or periodic structures~\cite{Koenderink2010,Sauvan2013,Raman2010}. The Green's function operator avoids these difficulties because it assumes a fixed frequency and is independent of material, and therefore of material loss. An additional advantage of the VIE approach is the compact domain, which sidesteps the subtle normalization~\cite{Snyder1983,Collin1991,Leung1994,Kristensen2012,Sauvan2013} required in the resonant-frequency approach, where the ``leaky'' fields diverge as they extend to infinity. 

We consider open systems---in which energy can enter and exit---such that typical operators, including the curl-curl operator $\mathcal{M}$ of \eqref{Maxwell_Eig} and the Green's function integral operator $\mathcal{G}$ of \eqref{VIE}, are not Hermitian. By reciprocity $\mathcal{G}$ is complex-symmetric~\cite{Novotny2012}, such that if it is diagonalizable, its generic $\mathcal{U} \Xi \mathcal{U}^{-1}$ eigendecomposition can be written~\cite{Horn1986}
\begin{align}
    \mathcal{G} = \mathcal{U} \Xi \mathcal{U}^{T},
    \label{eq:GDiag}
\end{align}
where $\Xi$ is a diagonal operator with entries $\xi_n$ and $\mathcal{U}$ is the basis of eigenfunctions $\vect{e}_n$. The assumption of diagonalizability (the existence of a ``spectral'' eigendecomposition of the operator) is commonplace is physics. Technically, even for Hermitian operators, rigorous eigendecomposition in infinite-dimensional spaces is subtle and subject to obscure counter-examples~\cite{Reed1980,Pedersen1989}. However, if one makes the reasonable conjecture that the system can be simulated on a computer, i.e. that one has a convergent finite-dimensional discretization, then one should be able to apply the eigendecomposition in this finite-dimensional system to arbitrary accuracy; this can be viewed as a justification of the commonplace assumption of diagonalizability. Even in a finite-dimensional problem, of course, diagonalizability of a non-Hermitian matrix is not guaranteed~\cite{Horn1986}, but breakdown of diagonalizability only occurs at ``exceptional points'' that must be forced~\cite{Kato1980,Heiss2004}. Exceptional points occur by chance with zero probability; near an exceptional point, the operator $\mathcal{G}$ is highly nonnormal~\cite{Trefethen2005} and ill-conditioned (i.e. ${\rm cond}(\mathcal{U}\mathcal{U}^\dagger) \gg 1$), but it is diagonalizable, and we show in the text that the response of such structures cannot surpass our limits. By continuity our limits also apply at non-diagonalizable exceptional points, thereby justifying our assumption of diagonalizability. Our decomposition is similar in spirit to the singularity and eigenmode expansion methods, SEM and EEM, respectively, for surface integral equations in electromagnetic scattering ~\cite{Baum1991,Baum1976,Rothwell2001,Ramm1980,Dolph1980}.

The complex-symmetry of the operator leads to an atypical (unconjugated, indefinite) ``inner product'' under which the eigenfunctions are orthogonal: $\vect{e}_i^T \vect{e}_j = \int_V \vect{E}_i \cdot \vect{E}_j = \delta_{ij}$.  Because modes are not orthogonal under the typical inner product $\vect{e}^\dagger \vect{e} = \int_V \cc{\vect{E}} \cdot \vect{E}$, they are not \emph{power}-orthogonal~\cite{Collin1991}: it is possible for energy in one mode to mix into another, leading to effects such as destructive interference in scattering ``dark states''~\cite{Hsu2014} (e.g. Fano resonances~\cite{Fano1961}) and the Petermann factor for noise enhancement in lasers~\cite{Petermann1979,Siegman1989}. Another possibility is that $\int_V \vect{E}_i \cdot \vect{E}_i = 0$, i.e. ``self-orthogonality,'' which renders the eigenfunction basis incomplete~\cite{Brink2001} and corresponds to exceptional points~\cite{Heiss2004,Liertzer2012}. As shown in the text, energy conservation prevents any of these phenomena from surpassing our limits.

A modal decomposition of the electric field $\vect{e}$ follows from the modal decomposition of $\mathcal{G}$ in \eqref{GDiag}. In order to isolate $\mathcal{G}$, it is easier to work with the polarization currents, $\chi \vect{e}$, which from \eqref{VIE} are given by:
\begin{align}
    \chi \vect{e} &= \chi(\omega) \left(\mathcal{I} + \chi(\omega) \mathcal{G}\right)^{-1} \vect{e}_{\rm inc} \nonumber \\
                  &= \mathcal{U} \left(\Xi - \xi(\omega)\right)^{-1} \mathcal{U}^T \vect{e}_{\rm inc},
    \label{eq:modalE}
\end{align}
where $\xi(\omega) = -1/\chi(\omega)$ and the inverse of $\mathcal{I} + \chi \mathcal{G}$ is guaranteed to exist (for $\varepsilon \neq 0,1$) because it is a Fredholm operator~\cite{Costabel2012}.

Because the extinguished power and the total LDOS are both imaginary parts of linear functionals of the induced fields, their modal decompositions are single sums over the resonances. Given the electric field from \eqref{modalE}, the extinction of \eqref{pExt} and total LDOS of \eqref{rhoTot} are determined by overlap integrals between the incident field and the VIE-basis eigenfunctions:
\begin{subequations}
\begin{align}
    P_{\rm ext} &= \frac{\varepsilon_0 \omega}{2} \Im \left[ \vect{e}_{\rm inc}^\dagger \mathcal{U} \left(\Xi - \xi(\omega)\right)^{-1} \mathcal{U}^T \vect{e}_{\rm inc} \right] \label{eq:extModal} \\ 
    \frac{\rho_{\rm tot}}{\rho_0} &= 1 + \frac{2\pi n_0 \omega}{c} \Im \left[ \vect{e}_{\rm inc}^T \mathcal{U} \left(\Xi - \xi(\omega)\right)^{-1} \mathcal{U}^T \vect{e}_{\rm inc} \right] \label{eq:rhoModal}.
\end{align}
\end{subequations}
Because $\left(\Xi - \xi(\omega)\right)^{-1}$ is a diagonal operator, it can be written as a single sum over the modes, simplifying the extinction and LDOS: 
\begin{subequations}
\begin{align}
    P_{\rm ext} &= \frac{\varepsilon_0 \omega}{2} \Im \sum_n \frac{p_n}{\xi_n -
\xi(\omega)} \label{eq:Pext_sum} \\
    \frac{\rho_{\rm tot}}{\rho_0} &= 1 + \frac{2\pi}{k^3} \Im \sum_n \frac{\rho_n}{\xi_n -
\xi(\omega)} \label{eq:rho_sum}
\end{align}
\end{subequations}
where $p_n = \int_V \cc{\vect{E}}_{\rm inc} \cdot \vect{E}_n \int_V \vect{E}_n \cdot \vect{E}_{\rm inc}$, $\rho_n = \sum_j \left(\int_V \vect{E}_n \cdot \vect{E}_{\textrm{inc},s_j}\right)^2$, and $\xi_n$ are the VIE eigenvalues and the diagonal entries of $\Xi$. The $p_n$ and $\rho_n$ are normalized ``oscillator strengths'' representing the per-mode extinguished power and per-mode total density of states, respectively. Adding up only the oscillator strengths yields structure-independent sum rules for each quantity:
\begin{subequations}
\begin{align}
    \sum_n p_n &= \int_V |\vect{E}_{\rm inc}|^2  \,{\rm d}V \label{eq:Psumrule} \\
    \sum_n \rho_n &= \sum_j \int_V \vect{E}_{\textrm{inc},s_j} \cdot \vect{E}_{\textrm{inc},s_j} \,{\rm d}V,
    \label{eq:rhosumrule}
\end{align}
\end{subequations}
where the sum over all modes exploits the eigenbasis orthogonality, $\mathcal{U}^T \mathcal{U} = \mathcal{I}$, and as before $j$ indexes the polarization of the dipole emitter. For each quantity, the sum of the oscillator strengths $p_n$ or $\rho_n$ is given by the intensity of the electric field originally incident upon the volume occupied by the scatterer.

Ideally, the sum rules for the extinction and the total LDOS would lead directly to limits, with the numerators in \eqreftwo{Pext_sum}{rho_sum} bounded above by the sum rules and the denominators bounded below by $\Im \xi(\omega) = \Im \chi(\omega) / |\chi(\omega)|^2$, which is nonzero due to material losses. Because our system is non-Hermitian, such an argument is not valid. If we had a Hermitian system with a conjugated orthogonality relationship $\int \cc{\vect{E}}_i \cdot \vect{E}_j = 0$, then we would have used $\mathcal{U}^\dagger$ instead of $\mathcal{U}^T$ in \eqref{GDiag} and the resulting amplitudes $p_n$ and $\rho_n$ would have been real and positive. Due to radiation losses (though \emph{not} metal losses), we have complex $p_n$ and $\rho_n$ with possibly negative real parts, and hence it is possible to have e.g. $|p_n| \gg |\sum p_n|$. Such a response is a general feature of \emph{nonnormal} dynamics, where there can be significant amplification beyond what one would expect from the resonances, as the pseudospectral level curves~\cite{Trefethen2005} may look very different from typical circles centered at the eigenvalues.

Energy conservation prevents such responses from surpassing the sum rule limits that are obtained when all of the oscillator strength in \eqreftwo{Psumrule}{rhosumrule} are concentrated at a single resonance (with the caveat that $\vect{E}_{\textrm{inc},s_j} \rightarrow \cc{\vect{E}}_{\textrm{inc},s_j}$, although for the dominant $1/r^3$ quasistatic term this makes no difference). The optimal resonance is given by 
\begin{align}
    \xi_{\rm ext,opt} &= \Re \xi(\omega)
\end{align}
which is on the real line, as close to the material parameter $\xi(\omega)$ as possible. This choice of eigenvalue leads to the extinction limit in \eqref{optAbsSigmaMetal}. Imposing energy conservation on the absorption, scattering, and radiative and nonradiative LDOS integrals in the VIE approach yields the limits of Eqs.~(\ref{eq:optScatSigmaMetal},\ref{eq:optAbsSigmaMetal},\ref{eq:rhoRadLimitMetal},\ref{eq:rhoNrLimitMetal}), but we will not prove that here.

An interesting possibility that arises from the VIE approach is the potential existence of further sum rules. In addition to considering the sum of oscillator strengths in \eqref{Psumrule}, one can consider the sum of eigenvalues, weighted by the oscillator strengths:
\begin{align}
    \sum_n &\xi_n p_n \nonumber \\
    &= \sum_n \int\int \vect{E}^{\rm inc}_i(\vect{x}) \vect{E}^n_i(\vect{x}) \xi_n \vect{E}^n_j(\vect{x}') \vect{E}^{\rm inc}_j(\vect{x}') \nonumber \\
                     &= \sum_n \int\int\int \vect{E}^{\rm inc}_i(\vect{x}) \vect{E}^n_i(\vect{x}) \tens{G}_{jk}(\vect{x}',\vect{x}'') \vect{E}^n_k(\vect{x}'') \vect{E}^{\rm inc}_j(\vect{x}') \nonumber \\
                     &= \int\int \vect{E}^{\rm inc}_i(\vect{x}) \tens{G}_{ji}(\vect{x}',\vect{x}) \vect{E}^{\rm inc}_j(\vect{x}')
    \label{eq:eig_sum_rule}
\end{align}
where we used the resolution of the indentity $\sum_n \vect{E}^n_i(\vect{x}) \vect{E}^n_k(\vect{x}'') = \delta_{ik} \delta(\vect{x} - \vect{x}'')$ to simplify the third line. In the surface-integral representation of quasistatic electromagnetism it can be shown the oscillator strengths and relevant eigenvalues for extinction averaged over all angles is constrained to satisfy $\sum_n \xi_n p_n / \sum_n p_n = 1/3$, reducing the possibility for all-angle response relative to the single-angle response~\cite{Miller2014,Fuchs1976}. \Eqref{eig_sum_rule} and its higher-order counterparts (e.g. $\sum_n \xi_n^2 p_n \sim \vect{e}_{\rm inc}^\dagger \mathcal{G}\mathcal{G} \vect{e}_{\rm inc}$) may yield stricter sum rules under various incident fields, reducing the possible response.

\section{Bounds on the $\BigO{kL}$ term in the LDOS limits}
\label{app:vol_integrals}
The bounds on the radiative and nonradiative LDOS, \eqreftwo{rhoRadLimitMetal}{rhoNrLimitMetal}, take into account the $1/r^3$, $1/r^2$, and $1/r$ terms in the free-space dyadic Green's function. Integrating the $1/r$ term over a half-space (or a spherical shell, or any other structure separated some distance $d$ from the source) yields the $\BigO{kL}$ term in \eqreftwo{rhoRadLimitMetal}{rhoNrLimitMetal}, which diverges as the size $L\rightarrow \infty$. As discussed in the text, this divergence is unphysical. It results from deriving the optimal current as proportional to the incident field, $\vect{P} \sim \vect{E}_{\rm inc}$, which is appropriate and feasible for the evanescent waves, but which for the $1/r$ term yields a physically impossible fixed energy density over infinite space within a lossy medium. 

Despite the divergence of this term, for \emph{finite} object sizes its contribution to the limits of \eqreftwo{rhoRadLimitMetal}{rhoNrLimitMetal} is actually very small. One does not even have to consider a finite object but a finite interaction distance: $L$ represents the largest distance over which polarization currents in the metal generate nonzero scattered fields at the dipole source. Even wavelength-scale lengths are upper bounds to reasonable interaction
distances in a lossy medium. Table~1 compares the near-field limit in \eqref{rhoRadLimitMetalNF} to the full limit with all terms in \eqref{rhoRadLimitMetal} as well as the full limit without the $\BigO{kL}$ term, 
\begin{align}
    \frac{\rho_{\rm nr}}{\rho_0} \leq \frac{\left|\chi\right|^2}{\Im \chi} \left[\frac{1}{32(kd)^3} + \frac{1}{16 kd}\right] + 1. 
    \label{eq:rhoRadNoDiv}
\end{align}
We take $|\chi|^2/\Im \chi=1$ for simplicity. One can see from Table~1 that the near-field limit is a very good approximation to the overall limit for realistic separations, and that the far-field term contributes at most $0.03\%$ for the cases considered.

\begin{table}[h]
\begin{center}
\caption{Tabulation of higher-order term contributions to radiative LDOS limits.}
\begin{tabular}[c]{|c|c|c|c|c|c|c|}
    \hline
    kd & kL & \eqref{rhoRadLimitMetalNF} & \eqref{rhoRadNoDiv} & Rel. Error & \eqref{rhoRadLimitMetal} & Rel. Error \\ \hline
    0.0628 & 1 & 125.98 & 127.97 & 1.56$\%$ & 128.02 & +0.03$\%$ \\ \hline
    0.01 & 1 & 31250 & 31257.25 & 0.0232$\%$ & 31257.3 & +0.0001$\%$ \\ \hline
    0.01 & 10 & 31250 & 31257.25 & 0.0232$\%$ & 31257.68 & +0.0013$\%$ \\ \hline
\end{tabular}
\end{center}
\end{table}

\section{Suppressing absorption}
\label{app:supp_abs}
As discussed in the text, the optimal scattering limits are reached when absorption and scattering are exactly equal. For some applications (e.g. solar cells enhanced by plasmonic particle scatterers~\cite{Atwater2010}), however, parasitic absorption is detrimental and needs to be avoided, even if the per-volume scattering is reduced. Here we present alternate limits to \eqref{optScatSigmaMetal} to account for absorption suppression.

We define a fraction $f$ that is the ratio of absorption to extinction, 
\begin{align}
    f = \frac{P_{\rm abs}}{P_{\rm abs} + P_{\rm scat}}.
\end{align}
Suppose we define our figure of merit as the maximum scattering cross-section per unit volume subject to the condition that $f$ is smaller than some maximum ratio $f_{\rm max}$:
\begin{align}
    \max \quad&P_{\rm scat} \nonumber \\
    \textrm{s.t.} \quad&f < f_{\rm max}. \nonumber
\end{align}
As in \secref{limits}, we use standard Lagrangian optimization techniques and the the fact that the constraint on $f$ is active ($f = f_{\rm max}$) by the KKT conditions. By the same steps as in \secref{limits}, it is straightforward to show that the per-volume scattering cross-section is limited by:
\begin{subequations}
\begin{align}
    \frac{\sigma_{\rm scat}}{V} &\leq f_{\rm max}\left(1-f_{\rm max}\right) k \frac{|\chi(\omega)|^2}{\Im\chi(\omega)}.
    \label{eq:optScatSigmaApp}
\end{align}
\end{subequations}
\eqref{optScatSigmaApp} is maximized at $f_{\rm max} = 0.5$, corresponding to the unconstrained optimum in \eqref{optScatSigma}. For significantly reduced absorption (say $<2\%$), there is a significant penalty ($>10$$\times$) to the maximum per-volume scattering.

\section{Quasistatic cross-section of a coated sphere}
\label{app:coated_sphere_sigma}
Also shown in \figref{extinction_structures} are the cross-sections of coated spheres that have been optimized by the same procedure as the ellipsoids. Coated spheres, with dielectric cores and metallic shells, are a common structure for tunable resonances across visible and infrared frequencies~\cite{Averitt1999,Dreaden2012,Loo2005,Lal2007,Jain2006}, but one can see that their response falls short of the limits. To understand why the performance falls short, we consider the quasistatic absorption cross-section, which is known analytically~\cite{Bohren1983} and can be compared to \eqref{sv_ell_qs}. For simplicity we assume the particle core has the same permittivity as the shell, which has only a small effect on the response but enables us to write the typical~\cite{Bohren1983} coated-sphere cross-section $\sigma_{\rm cs}$ in the form
\begin{align}
    \left(\frac{\sigma_{\rm abs}}{V}\right)_{\textrm{cs}} = \frac{f_V k}{L_1 - L_0} \Im \left[ \frac{2/3 - L_0}{L_0 - \xi(\omega)} + \frac{L_1 - 2/3}{L_1 - \xi(\omega)}\right]
\end{align}
where $f_V$ is the metal volume fraction, $L_0 = 1/2 - 1/6\sqrt{1+8f_V}$ and $L_1 = 1/2 + 1/6\sqrt{1+8f_V}$ are structural depolarization factors, and as in \secref{limits} we define $\xi(\omega) = -1/\chi(\omega)$. One can see that there are two quasistatic resonances that arise from coupling the plasmons at the interior and exterior interfaces. At visible and infrared frequencies, the material parameters of typical metals satisfy $|\Re \chi| \gg 1$, such that a single resonance dominates the response and the volume fraction $f_V$ of the metal must be small (as in \figref{extinction_structures}). The cross-section is then given by
\begin{align}
    \left(\frac{\sigma_{\rm abs}}{V}\right)_{\textrm{cs}} \approx \frac{2}{3} f_V k \frac{1}{\Im \xi(\omega)} \approx \frac{2}{3}k \frac{|\Re \chi(\omega)|}{\Im \chi(\omega)}
\end{align}
where we see that the absorption cross-section is proportional to $|\Re \chi| / \Im \chi$ instead of $|\chi|^2 / \Im \chi$ due to the small optimal volume fraction, which decreases as $\Re \chi$ increases. There is an additional 33$\%$ reduction due to the second ``material'' resonance~\cite{Miller2014} at $L_1$. Additionally, tuning the resonant wavelengths requires very large core-shell thickness ratios, roughly 25:1 at $\lambda=1000$nm, whereas at the same wavelength the optimal ellipsoids have aspect ratios of less than 10:1.

\section{Propagation length of a plasmon near the surface-plasmon frequency}
\label{app:plasmon_prop_length}
It is sometimes stated~\cite{Raether1988,Barnes2003,Dionne2005,Barnes2006,Homola2006} that for any frequency at which $|\Re \varepsilon_m| \gg \Im \varepsilon_m$, for metallic permittivity $\varepsilon_m$, the complex surface-parallel wavevector for a plasmon at a metal--dielectric interface is given by:
\begin{align}
    \beta = \frac{\omega}{c} \left[ \sqrt{\frac{\varepsilon_m' \varepsilon_d}{\varepsilon_m' + \varepsilon_d}} + i\frac{\varepsilon_m''}{2\left(\varepsilon_m'\right)^2} \left(\frac{\varepsilon_m' \varepsilon_d}{\varepsilon_m' + \varepsilon_d}\right)^{3/2} \right],
    \label{eq:approx_beta}
\end{align}
where $\varepsilon_d$ is the dielectric permittivity and we use single and double primes to denote real and imaginary parts, resp. However, this expression is only an approximation, and it is only accurate at low frequencies where the plasmon confinement to the surface is \emph{not} highly subwavelength. In particular, it is not accurate near the resonant frequency at which $\varepsilon_m' \approx -\varepsilon_d$, even if $\left|\varepsilon_m'\right| \gg \varepsilon_m''$. The frequency-dependent propagation length of the plasmon is given by:
\begin{align}
    L_{\rm prop} = \frac{1}{2 \Im \beta},
\end{align}
which according to \eqref{approx_beta} would go to zero at $\varepsilon_m' = -\varepsilon_d$, even for a lossy medium (as is claimed in \citeasnoun{Dionne2005}). This is not correct.

The exact dispersion relation for a surface plasmon is~\cite{Raether1988}
\begin{align}
    \beta = \frac{\omega}{c} \sqrt{\frac{\varepsilon_m \varepsilon_d}{\varepsilon_m + \varepsilon_d}}.
    \label{eq:full_beta}
\end{align}
The approximation in \eqref{approx_beta} arises from a Taylor expansion of the imaginary part of $\varepsilon_m$ in the denominator  of \eqref{full_beta}. However, such a Taylor expansion is only valid if $|\varepsilon_m' + \varepsilon_d| \gg \varepsilon_m''$. Thus we see that this approximation fails near the surface-plasmon frequency, which is the exact regime where significant subwavelength confinement is possible. If we define the surface plasmon frequency $\omega_{\rm sp}$ such that $\varepsilon_m'(\omega_{\rm sp}) = -\varepsilon_d$, then the dispersion relation is approximately
\begin{align}
    \beta \left(\omega_{\rm sp}\right) \approx \frac{\omega}{c} \frac{\varepsilon_d}{\sqrt{\varepsilon_m''}} \left(\frac{1+i}{\sqrt{2}}\right),
\end{align}
assuming $\left|\varepsilon_m'\right| \gg \varepsilon_m''$ (as above). It is then straightforward to derive the propagation length at $\omega = \omega_{\rm sp}$:
\begin{align}
    L_{\rm prop} \approx \frac{\lambda}{2\sqrt{2}\pi} \frac{\sqrt{\varepsilon_m''}}{\varepsilon_d},
\end{align}
which does not go to zero except in the case of zero loss. (In the case of zero loss the group velocity also goes to zero, with the same dependence on $\varepsilon_m''$, such that the decay time is nonzero and finite. As the material loss goes to zero, the energy loss per unit time remains nonzero but finite, because the resonant field enhancement balances the vanishing material loss.) Thus near the surface-plasmon frequency the propagation length is proportional to $\sqrt{\Im \varepsilon_m}$, not $\left(\Re \varepsilon\right)^2 / \Im \varepsilon$ (which is the scaling of \eqref{approx_beta} and which only applies in the low-frequency Sommerfeld--Zenneck regime).

\section{Optimal LDOS of a thin film}
\label{app:thin_film_ldos}
We derive \eqref{ldos_thin_film}, the optimal LDOS of a thin film at frequencies $\omega \ll \omega_{\rm sp}$. We measure the LDOS a distance $d$ from the thin film, which we take to have a thickness $t$. The LDOS of a thin film, relative to the free-space LDOS $\rho_0$, is~\cite{Joulain2005}
\begin{align}
    \frac{\rho}{\rho_0} \approx \frac{1}{k^3} \int k_p^2 e^{-2 k_p z} \left(\Im S_{21}\right) \,{\rm d}k_p
\end{align}
where $k_p$ is the magnitude of the surface-parallel wavevector, and we have assumed the primary contribution to the integral comes from high-wavevector (p-polarized) waves for which $e^{2i k_z d} \approx e^{-2 k_p d}$. For small imaginary permittivity $\Im \varepsilon \ll \left|\Re \varepsilon\right|$, which is a requirement to access the large wavevectors, $\Im S_{21}$ is a sharply peaked function centered around a parallel wavevector $k_{p0}$ that depends on the thin-film thickness $t$ by~\cite{Miller2014a}
\begin{align}
    k_{p0} \approx \frac{2}{\left|\varepsilon\right| d}
\end{align}
which we will use below to determine the optimal thickness. The integral can be approximated by
\begin{align}
    \int k_p^2 e^{-2 k_p z} \left(\Im S_{21}\right) \,{\rm d}k_p \approx \frac{\pi}{2} k_{p0}^2 e^{-2 k_{p0} z} \left[\Im S_{21}\right]_{\rm max} \Delta k_p 
\end{align}
where we have assumed that $\Im S_{21}$ is a Lorentzian with full-width half-max of $\Delta k_p$.  As derived in \citeasnoun{Miller2014a}, the ``reflectivity--bandwidth product'' is
\begin{align}
    \left[\Im S_{21}\right]_{\rm max} \Delta k_p \approx 2 k_{p0}
\end{align}
Thus the LDOS enhancement is proportional to $k_{p0}^3 e^{-2 k_{p0}z}$, which reaches a maximum for a thickness such that $k_{p0} = 3 / 2z$, yielding an optimal LDOS of
\begin{align}
    \left[\frac{\rho}{\rho_0}\right]_{\rm max} &\approx \frac{27 \pi}{8 e^3} \frac{1}{\left(kd\right)^3} \nonumber \\
                                               &\approx \frac{1}{2 \left(kd\right)^3}
\end{align}

\section{Frobenius norm of the dyadic Green's function}
\label{app:GF_frob_norm}
The dyadic Green's function defined in \eqref{GEP_frob_norm} is given by~\cite{Chew1995}
\begin{align}
    \tens{G^{EP}} = \frac{k^2 e^{ikr}}{4\pi r} \left[\left(1 + \frac{i}{a} - \frac{1}{a^2}\right)\delta_{ij} + \left(-1 - \frac{3i}{a} + \frac{3}{a^2}\right) \frac{x_i x_j}{r^2}\right]
\end{align}
where $a=kr$. The Frobenius norm is given by $G^{EP}_{ij}\overline{G^{EP}_{ij}}$ (momentarily denoting complex conjugation with an overline) and can be computed by use of the identities (repeated indices indicating summation)
\begin{align*}
    &\delta_{ij} \delta_{ij} = \delta_{ii} = 3 \\
    &\delta_{ij} \frac{x_i x_j}{r^2} = \frac{x_i x_i}{r^2} = 1 \\
    &\frac{x_i x_j}{r^2} \frac{x_i x_j}{r^2} = 1
\end{align*}
which yield the expression in \eqref{GEP_frob_norm}.

\section*{Acknowledgments}
We thank Eli Yablonovitch, Adi Pick, Yehuda Avniel, Aristeidis Karalis, Hrvoje Buljan, Alejandro Rodriguez, and Emma Anquillare for helpful discussions. This work was supported by the Army Research Office through the Institute for Soldier Nanotechnologies under Contract No. W911NF-07-D0004, and by the AFOSR Multidisciplinary Research Program of the University Research Initiative (MURI) for Complex and Robust On-chip Nanophotonics under Grant No. FA9550-09-1-0704.

\end{document}